\documentclass[useAMS,usenatbib,referee]{mn2e}
\usepackage{txfonts,graphicx,natbib,color,verbatim}
\bibpunct{(}{)}{;}{a}{}{,}

\def\lesssim{\mathrel{\hbox{\rlap{\hbox{\lower4pt\hbox{$\sim$}}}\hbox{$<$}}}}
\def\gtrsim{\mathrel{\hbox{\rlap{\hbox{\lower4pt\hbox{$\sim$}}}\hbox{$>$}}}}

\newcommand{\msol}{\mbox{M$_{\odot}$}}

\newcommand{\logl}{\mbox{$\log L/{\rm L_{\odot}}$}}

\DeclareMathAlphabet{\mathsc}{OT1}{cmr}{m}{sc}
\def\testbx{bx}%
\DeclareRobustCommand{\ion}[2]{%
\relax\ifmmode
\ifx\testbx\f@series
{\mathbf{#1\,\mathsc{#2}}}\else
{\mathrm{#1\,\mathsc{#2}}}\fi
\else\textup{#1\,{\mdseries\textsc{#2}}}%
\fi}

\title[The progenitors of SNe 2006my, 2006ov \& 2004et]
      {On the nature of the progenitors of three type II-P supernovae: 2004et, 2006my and 2006ov}
\author[R. M. Crockett {\rm et al.}]
{R.M. Crockett$^{1,2}$\thanks{E-mail: mark.crockett@astro.ox.ac.uk}, S.J. Smartt$^1$, A. Pastorello$^1$, J. J. Eldridge$^3$, A. W. Stephens$^4$, \newauthor J. R. Maund$^{5,6}$, S. Mattila$^7$\\
$^1$Astrophysics Research Centre, School of Mathematics and Physics, Queen's University Belfast, Belfast BT7 1NN, UK\\
$^2$Department of Physics, University of Oxford, Keble Road, Oxford, OX1 3RH, UK\\
$^3$Institute of Astronomy, University of Cambridge, Madingley Road, Cambridge CB3 0HA\\
$^4$Gemini Observatory, 670 North A'ohoku Place, Hilo, HI 96720, USA\\
$^5$DARK Cosmology Centre, Neils Bohr Institute, University of Copenhagen, Julian Maries Vej 30, 2100 Copenhagen, Denmark\\
$^6$Sophie and Tycho Brahe Fellow\\
$^7$Tuorla Observatory, Department of Physics and Astronomy, University of Turku, V$\ddot{a}$is$\ddot{a}$l$\ddot{a}$ntie 20, FI-21500 Piikki$\ddot{o}$, Finland.}

\date{Submitted 2010 Xxxxx XX}

\pagerange{\pageref{firstpage}--\pageref{lastpage}} \pubyear{2010}

\begin{document}
\bibliographystyle{aa}
\label{firstpage}

\maketitle

\begin{abstract}

The pre-explosion observations of the type II-P supernovae 2006my, 2006ov and 2004et, are re-analysed.  In the cases of supernovae 2006my and 2006ov we argue that the published candidate progenitors are not coincident with their respective supernova sites in pre-explosion Hubble Space Telescope observations.  We therefore derive upper luminosity and mass limits for the unseen progenitors of both these supernovae, assuming they are red supergiants: 2006my ($\log L/L_{\odot} = 4.51$; $m < 13$\msol ) and 2006ov ($\log L/L_{\odot} = 4.29$; $m < 10$\msol).  In the case of supernova 2004et we show that the $yellow~supergiant$ progenitor candidate, originally identified in Canada France Hawaii Telescope images, is still visible $\sim$3 years post-explosion in observations from the William Herschel Telescope.  High-resolution Hubble Space Telescope and Gemini (North) adaptive optics late-time imagery reveal that this source is not a single yellow supergiant star, but rather is resolved into at least three distinct sources.  We report the discovery of the unresolved progenitor as an excess of flux in pre-explosion Isaac Newton Telescope i'-band imaging.  Accounting for the late-time contribution of the supernova using published optical spectra, we calculate the progenitor photometry as the difference between the pre- and post-explosion, ground-based observations.  We find the progenitor was most likely a late K to late M-type supergiant of $8^{+5}_{-1}$ M$_{\odot}$.  In all cases we conclude that future, high-resolution observations of the supernova sites will be required to confirm these results.

\end{abstract}


\section{Introduction}

\setlength\parindent{4mm}

Type II-plateau (II-P) supernovae (SNe) have long been held to be the core-collapse induced explosions of red supergiant stars - evolved massive stars with cool ($\sim$3000\,K), expansive ($\sim$10$^3$\,R$_{\odot}$) hydrogen (H) rich envelopes \citep[for a detailed review see][]{2009ARA&A..47...63S}.  That the progenitors of such SNe should be H-rich follows from the observation of strong H lines in the spectra of these events, the definition of a type II SN \citep[see for example,][]{Filippenko:1997p259}.  For a star to undergo core-collapse it must be of sufficient initial mass in order to proceed to the ultimate stages of nuclear burning and thereby attain an iron core.  Unable to counteract gravity through further nuclear fusion, the iron core contracts, but is supported against complete collapse by electron degeneracy pressure.  When the core reaches the Chandrasekhar mass limit ($\sim$1.4\,M$_{\odot}$), electron degeneracy pressure is overcome and the core rapidly collapses until neutron degeneracy pressure brings it to a halt, forming a neutron star.  This collapse releases around 10$^{53}$ erg of gravitational potential energy, which flows outwards from the collapsed core in the form of neutrinos.  Roughly 1 percent of this neutrino flux (10$^{51}$ erg) couples with the outer layers of the star, causing the supernova explosion.  For a review of the core-collapse process see, for example, \citealt{1990RvMP...62..801B}.  Theory tells us that the initial mass threshold for a star to produce a collapsing core is between 7-12\,M$_{\odot}$ \citep{Heger:2003p40,Eldridge:2004p60,Siess:2007p64,Poelarends:2008p103}.

Attempts to directly observe SN progenitors are hampered by a range of issues: the rarity of SN explosions ($\sim$1 SN per galaxy per 100 years); the difficulty in resolving single stars in galaxies outside of our own; and the lack of deep, high-quality pre-explosion observations of SN sites.  Not surprisingly few SNe overcome these problems, but in recent years the archive of the Hubble Space Telescope (HST) has gone some way to increase the numbers that do.  The diffraction limited view of HST has made it possible to identify single massive stars in galaxies at distances of up to $\sim$ 20 Mpc.  

For the last decade, several research groups around the world have been systematically searching for the progenitors of Core-Collapse SNe in archival imagery, for the most part in data from HST \citep[e.g.,][]{2003PASP..115..448V,VanDyk:2003p1940,2001ApJ...556L..29S,Smartt:2003p1939,Li:2005p1115,Li:2007p656,2008ApJ...681L...9P,2009MNRAS.398.1041B}.  It was in archival HST and high quality Gemini telescope imagery that the first confirmed discovery of a red supergiant progenitor for a type II-P SN was made; that of SN 2003gd, with an estimated mass of $\sim$8\,M$_{\odot}$ \citep{Smartt:2004p444,VanDyk:2003p487,2009Sci...324..486M}.  This was followed by the detection of the red supergiant progenitor of SN 2005cs \citep{Maund:2005p490,Li:2006p1886,Eldridge:2007p556}, again in archival HST data.  Several other SN II-P progenitor candidates have been reported in the literature: SN 1999ev \citep{Maund:2005p492,VanDyk:2003p1940}, SN 2004A \citep{Hendry:2005p494}, SN 2004et \citep{Li:2005p1115}, SN 2006my and SN 2006ov \citep{Li:2007p656}, SN 2008bk \citep{2008ApJ...688L..91M}, SN 2008cn \citep{2009ApJ...706.1174E}, and SN 2009kr \citep{2010ApJ...714L.280F}.  All were identified in HST data, with the exception of the progenitor of SN 2008bk which was found in images from the VLT.  More often than not, even where HST data were available, non-detections have been the norm.  In these cases, qualified upper luminosity and mass limits have been set from the depth of the pre-explosion images, the main caveat being the assumption that the progenitor was a red supergiant \citep[e.g.][]{Smartt:2003p1939, VanDyk:2003p1940}.

In an extensive paper \citep{2009MNRAS.395.1409S} we have attempted to place constraints on the progenitor population of type II-P SNe by examining the progenitor detections and detection limits of 20 such events in a homogeneous fashion.  From a maximum likelihood analysis of these mass limits assuming a Salpeter initial mass function (IMF), we found minimum and maximum initial masses for the progenitors of type II-P SNe of $m_{min} = 8.5^{+1}_{-1.5}$M$_{\odot}$ and $m_{max} = 16.5\pm1.5$M$_{\odot}$.   The minimum mass is in good agreement with that predicted by theory, but the maximum mass is lower than expected.  Stars with main-sequence masses of up to $\sim25-30$M$_{\odot}$ are expected to retain most of their H-rich envelope, becoming red supergiants and exploding as SNe II-P.  In \citet{2009MNRAS.395.1409S} we estimate that, if our 20 progenitors were actually sampled from a population with masses between 8.5 and 25M$_{\odot}$, we should have detected 4 progenitors of between 17-25M$_{\odot}$.  That we detected none by chance has a probability of just 2 per cent.  We have termed this 2.4$\sigma$ result the {\em red supergiant problem} and refer the reader to \citet{2009MNRAS.395.1409S} for detailed discussion.  For several of our sample of 20 objects in that paper we relied upon previously unpublished works by our group.  In this paper we report our detailed analysis of the pre-explosion observations of three of these type II-P supernovae: SNe 2006my, 2006ov and 2004et.

SN 2006my was discovered by K. Itagaki \citep{Nakano:2006p559} on 2006 Nov 8.82\,UT (all times reported here are UT), located at $\mathrm{\alpha_{J2000}=12^{h}43^{m}40^{s}.74,\delta_{J2000}=+16\degr23\arcmin14\arcsec.1}$, some 27$\arcsec$ west and 22.5$\arcsec$ south of the centre of its host galaxy NGC 4651.  \citet{Stanishev:2006p1929} spectroscopically classified SN 2006my as type II-P, similar to the class prototype SN 1999em \citep{Elmhamdi:2003p1943,Hamuy:2001p1942,Leonard:2002p1941} at 1-2 months past maximum.  \citet{Li:2007p656} have since shown that its discovery was probably $\sim$3 months after explosion by comparing its light curve with that of SN 1999em, and its spectra with with those of SN 2004dj.

SN 2006ov was discovered by K. Itagaki \citep{Nakano:2006p1928} on 2006 Nov 24.86 in M61 (NGC~4303), located at $\mathrm{\alpha_{J2000}=12^{h}21^{m}55^{s}.30,\delta_{J2000}=+4\degr29\arcmin16\arcsec.7}$, which is 5.5$\arcsec$ east and 51$\arcsec$ north of its host galaxy centre.  From a spectrum obtained on Nov 25.56, \citet{Blondin:2006p705} classified SN 2006ov as a type II, similar to the type II-P SN 2005cs around 1 month post-explosion although somewhat reddened in comparison.  However, they do not specify the degree of reddening.  Through detailed analysis of the SN light curve and spectra, \citet{Li:2007p656} show that SN~2006ov is not significantly reddened and was most probably discovered some 3 months post-explosion.  Li et al. point out that by mistakenly comparing its classification spectrum with one of SN 2005cs at around 1 month post-explosion, \citet{Blondin:2006p705} would have observed a much redder continuum for SN 2006ov and erroneously interpreted this as being due to extinction.  Since type II-P SNe become progressively redder as they evolve, the older age at discovery could explain its colour.

SN 2004et was discovered by S. Moretti \citep{Zwitter:2004p1930} on 2004 Sep 27, located at $\mathrm{\alpha_{J2000}=20^{h}35^{m}25^{s}.4,\delta_{J2000}=+60\degr07\arcmin17\arcsec.6}$ in the nearby spiral galaxy NGC~6946.  Spectroscopic and photometric monitoring revealed the presence of H Balmer lines in its optical spectra and a plateau lasting $\sim$110\,d in its early-time optical lightcurves, confirming it as a type II-P SN \citep{Li:2005p1115,Sahu:2006p1119,Misra:2007p1117,2010MNRAS.404..981M}.  Detection of the SN at X-ray and radio wavelengths \citep{Argo:2005p1792,Misra:2007p1117,MartiVidal:2007p1547} suggested the presence of significant circumstellar material (CSM), most likely the slow, dense wind of the progenitor. Interaction of the fast-moving SN ejecta with this material would create a hot, shocked region producing radio synchrotron and X-ray emission \citep{Chevalier:2006p1775}.  Furthermore, at about 3 years from the explosion the appearance of wide box shaped emission lines as well as rebrightening of the SN at mid-infrared wavelenths were interpreted as being due to the impact of the ejecta on the progenitor CSM (Kotak et al. 2009).  Extinction towards SN 2004et was estimated by \citet{Zwitter:2004p1930}, who measured the equivalent widths of the interstellar Na {\sc i} D lines from a high resolution spectrum of SN 2004et taken shortly after discovery, and calculated a total reddening (host + Galactic) of \mbox{E$(B\!-\!V)$} = 0.41, using the calibrations of \citet{Munari:1997p1098}.  It is to this value that the authors of the above refereed publications defer.

Detections of candidate progenitors have been reported for all three of these type II-P events - SNe~2006my and 2006ov by \citet{Li:2007p656} (hereafter Li07), and SN~2004et by \citet{Li:2005p1115} (hereafter Li05).  In $\S$\ref{sec:2006my} and $\S$\ref{sec:2006ov} we argue that the candidate progenitors of SNe 2006my and 2006ov are {\em not} coincident with their respective SN sites.  In $\S$\ref{sec:2004et}, our analysis shows that candidate, yellow-supergiant progenitor for SN 2004et is still visible some 3 years post-explosion.  This yellow source is actually a blend of several point sources, which are resolved in new HST and Gemini adaptive optics images.  We also present previously unpublished observations from the Isaac Newton Telescope (INT), in which we identify a red supergiant progenitor.

\section{The progenitor of SN 2006my}
\label{sec:2006my}

A candidate red supergiant
progenitor for SN~2006my was detected by Li07 in pre-explosion HST Wide Field and Planetary Camera 2 (WFPC2) observations (program GO-5375, PI: Rubin) of its host galaxy,
NGC4651.  They located the SN position on the HST data through alignment with ground based observations of the SN from the Canada-France-Hawaii Telescope (CFHT).  Thirteen sources common to both a CFHT+MegaCam $r'$-band SN image and the WFPC2 F814W observation were identified and their positions used to calculate a transformation between the respective coordinate frames.  The SN position was measured in the CFHT frame and transformed to the coordinate system of the archival HST data, yielding (x,y) coodinates of (410.61, 158.81) on WF2 of WFPC2 with a positional error of $\pm$0.45 pixels ($\pm$45 mas).  This error was defined as the rms error of the transformation.  Li07 identify a source within the error circle on the F814W image which they propose as the SN progenitor, characterising it as a single red supergiant star of main-sequence mass $M =10^{+5}_{-3}$M$_{\odot}$.

We have since obtained HST+WFPC2 observations of SN 2006my as part of our HST program GO-10803 (PI: Smartt).  The higher resolution of these data allow a more accurate alignment with the pre-explosion data to be determined.  The details of our analysis are reported below.  During our work, we became aware of a similar analysis using our HST follow-up observations being performed contemporaneously by \citet{Leonard:2008p1944}.  Both groups agreed that our analyses should proceed independently, and we have both arrived at the same conclusions.  Note that our basic result was originally published in \citet{2009MNRAS.395.1409S}, which references this paper in preparation.

\begin{figure*}
    \centering
    \includegraphics[width=160mm]{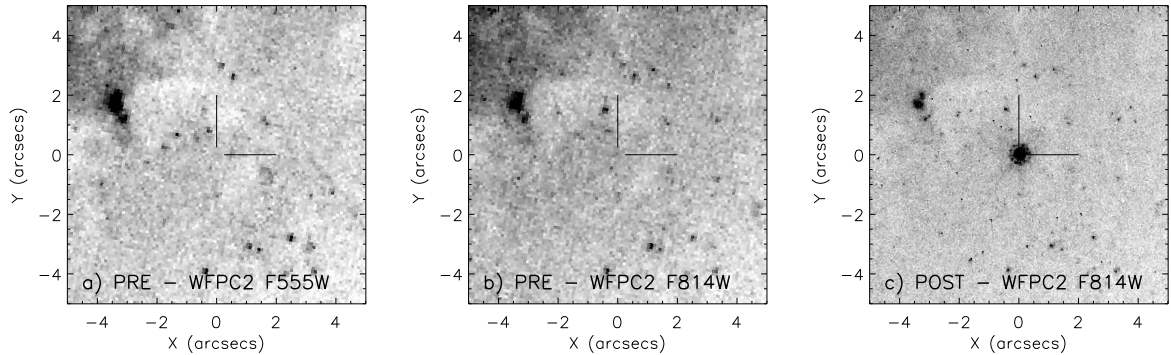}
    \caption{Aligned pre- and post-explosion HST+WFPC2 images of the field of SN 2006my.  All the images are centred on the SN position (marked by the cross hairs) and oriented such that North is up and East is to the left. {\bf (a)}: pre-explosion F555W, {\bf (b)}: pre-explosion F814W and {\bf (c)}: post-explosion F814W with the SN clearly visible.  A faint object is just visible close to the SN position in the pre-explosion F814W image.}
\label{fig:06my_fig}
\end{figure*}

\subsection{Astrometry and Photometry}
\label{sec:06my_ast_phot}

The pre- and post-explosion HST+WFPC2 observations of the site of SN~2006my were retrieved from the HST archive\footnote{http://archive.stsci.edu/hst/} at the Space Telescope Science Institute (STScI) via the on-the-fly recalibration (OTFR) pipeline (see Table~\ref{tab:06my_obstab}).  Each of the observations consisted of multiple exposures and these were co-added using the {\sc iraf stsdas} task {\em crrej} in order to remove cosmic rays.  The {\em crrej} task does not account for the significant geometric distortion suffered by WFPC2, which ranges from a few tenths of a pixel at the centre of each chip, to 2-3 pixels at the chip edges \citep{Gilmozzi_ISR,Holtzman:1995p1198, Casertano_ISR,2003PASP..115..113A}.  We have chosen to follow this reduction procedure (rather than employ the {\em Multidrizzle} image reconstruction pipeline which corrects for geometric distortion \citep{Fruchter:2002p1494,Multidrizzle_handbook}) in order to remain consistent with the reduction performed by Li07.  In this way we are able to quote measured positions that can be directly compared with those of Li07.  However, we go on to apply the distortion corrections derived by \citet{2003PASP..115..113A} and \citet{Kozhurina_ISR} before performing the alignment described in detail below.  To correct for the effects of geometric distortion on point source photometry, the {\em crrej} combined images were also multiplied by a geometric correction image\footnote{f1k1552bu - http://www.stsci.edu/hst/observatory/cdbs/SIfileInfo/WFPC2/WFPC2PixCorr}.

\begin{table}
\caption{Pre-and post-explosion observations of the site of SN 2006my.}
\begin{center}
\begin{tabular}{lcrc}
\hline\hline
Date  & Telescope+Instrument & Filter  & Exp. Time \\
      &            &         & (s)       \\
\hline
\multicolumn{4}{c}{{\bf Pre-explosion images}} \\
1994 May 20 & HST+WFPC2 & F555W & 660\\
1994 May 20 & HST+WFPC2 & F814W & 660\\
\\
\multicolumn{4}{c}{{\bf Post-explosion images}} \\
2007 Apr 27 & HST+WFPC2 & F450W & 1400\\
2007 Apr 26 & HST+WFPC2 & F555W & 1200\\
2007 Apr 27 & HST+WFPC2 & F814W & 1200\\
\hline\hline
\end{tabular}
\end{center}
\label{tab:06my_obstab}
\end{table}

It was noted that the pre-explosion F555W and F814W images were well aligned, hence a transformation between the pre- and post-F814W images only was used to determine the position of the SN on the pre-explosion frames.  The pre-explosion SN site was imaged on the WF2 chip of WFPC2 (pixel scale = 0.1$\arcsec$), while the SN was imaged on the PC chip (pixel scale = 0.0455$\arcsec$) in our follow-up observations.  Aided by the exquisite resolution of the HST data we identified 49 common point sources between the pre- and post-explosion F814W images.  The pixel positions were first corrected for geometric distortion using the chip and wavelength dependent transformations detailed in \citet{2003PASP..115..113A} and \citet{Kozhurina_ISR}.  \citet{2003PASP..115..113A} reports that these transformations are accurate to $\pm$0.01 pixels in the WF chips and $\pm$0.02 pixels in the PC chip.  The distortion corrected positions were subsequently used to derive a transformation between the "distortion-free" pre- and post-explosion coordinate frames.  The transformation (involving shifts, scales and rotations in x and y) was calculated using the {\sc iraf} task {\em geomap}, and found to have a total (x and y rms combined) rms scatter of $\pm$0.19 WF pixels (19 mas)\footnote{Unless otherwise stated, all coordinate uncertainties quoted it this paper are {\em total} values (i.e. x and y uncertainties combined in quadrature)}

To test the robustness of the fitted rms, we divided the 49 point sources into two lists, one of 24 and one of 25 sources, and calculated two independent transformations between the pre- and post-explosion frames, finding rms errors of $\pm$0.19 WF pixels for each fit.  Our aim was to test each transformation with point sources not used in its derivation.  The transformed positions of such objects are not affected by over-fitting that may have occurred in the {\em geomap} calculation, and as such can be used to derive an independent and robust estimate of the transformation rms. 

To that end, we used the coordinate transformation derived from one list to map the post-explosion positions from the opposite list onto the pre-explosion coordinate frame.  The fit rms was then calculated from the scatter of the {\em mapped} coordinates around the {\em measured} pre-explosion positions.  In this way we measured independent, rms errors of $\pm$0.20 WF pixels for each of the transformations.  These are consistent with the original rms values reported by {\em geomap} ($\pm$0.19 WF pixels), confirming there was little or no over-fitting (not surprising given that we employed only linear terms in our {\em geomap} fit) and that the geometric distortion correction was successful.  We adopt the slightly larger $\pm$0.20 WF pixel (20 mas) total rms for all of the above transformations.  (Details of all the transformation rms values are recorded in Table~\ref{tab:06my_rms}.)

\begin{table}
\caption{Fit residuals for {\sc iraf} {\em geomap} transformations between pre- and post-SN 2006my HST F814W observations.  See text for details.}
\begin{center}
\begin{tabular}{rcrr}
\hline\hline
49 source transform &\vline &24 source tranform & 25 source transform\\
$\sigma_x$ = 0.124 &\vline &$\sigma_x$ = 0.125 & $\sigma_x$ = 0.116 \\
$\sigma_y$ = 0.148 &\vline &$\sigma_y$ = 0.146 & $\sigma_y$ = 0.151 \\
$\sigma_{total}$ = 0.193 &\vline &$\sigma_{total}$ = 0.192 & $\sigma_{total}$ = 0.190 \vspace{3mm}\\
&\vline&\multicolumn{2}{l}{(residuals estimated from transformed positions}\\
&\vline&\multicolumn{2}{l}{of fit-independent source coordinates)}\\
&\vline &$\sigma_x$ = 0.134 & $\sigma_x$ = 0.140\\
&\vline &$\sigma_y$ = 0.151 & $\sigma_y$ = 0.145\\
&\vline &$\sigma_{total}$ = 0.202 & $\sigma_{total}$ = 0.202\\
\hline\hline
\multicolumn{3}{l}{All values are in units of WF chip pixels (0.1$\arcsec$ / pix)}&\\
\end{tabular}
\end{center}
\label{tab:06my_rms}
\end{table}

The position of SN~2006my in the post-explosion F814W frame was measured using the three centring algorithms of the {\sc daophot} task {\em phot} - centroid, Gaussian and ofilter - and the PSF-fitting photometry package HSTphot\footnote{Positions measured by HSTphot differ from those of {\sc daophot} by -0.5 in both x and y coordinates.  All positions quoted in this paper are corrected to {\sc daophot} values} (version 1.1.7b) \citep{Dolphin:2000p1123}.  The mean of the four measurements was adopted as the SN position, SN(x,y)$_{post}$ = (416.08, 448.54) $\pm$0.05 PC pixels (2.5 mas), estimated from the range of the four measurements.  This uncertainty is consistent with the limiting astrometric uncertainty of HSTphot \citep{Dolphin:2000p1123}.  

In order to map this SN position to the pre-explosion frame using our geomap transformations, we first had to correct for geometric distortion.  This was performed, as for the alignment point sources, using the distortion corrections detailed in \citet{2003PASP..115..113A} and \citet{Kozhurina_ISR} appropriate for the PC chip and the F814W filter.  The distortion corrected position was calculated as SN(x$_{\rm o}$,y$_{\rm o}$)$_{post}$ = (416.13, 448.68).

This was subsequently transformed to the pre-explosion coordinate system using the 49 point source transformation, yielding a (distortion corrected) progenitor position of SN(x$_{\rm o}$,y$_{\rm o}$)$_{pre}$ = (410.25, 159.21) on WF2, with a total uncertainty of $\pm$0.20 WF pixels (20 mas).  The transformation rms (20 mas) completely dominates over the contributions from the distortion correction (1 mas) \citep{2003PASP..115..113A} and the SN position measurement (2.5 mas).  (We also applied the 24 and 25 source transformations to the SN position, finding transformed coordinates within 0.01 WF pix (1 mas) of our value of SN(x$_{\rm o}$,y$_{\rm o}$)$_{pre}$ above.)

To compare our pre-explosion SN site position with that of Li07, we had to apply a geometric distortion correction (WF2, F814W correction from \citealt{Kozhurina_ISR}) to the Li07 position, which was quoted for the distorted coordinate frame.  The original Li07 SN site position (Li(x,y)$_{pre}$ = (410.61, 158.81) $\pm$0.45 WF pixels) becomes Li(x$_{\rm o}$,y$_{\rm o}$)$_{pre}$ = (410.62, 158.56).  Comparing with our value of SN(x$_{\rm o}$,y$_{\rm o}$)$_{pre}$ = (410.25, 159.21) we find an offset of 0.75$\pm$0.49 WF pixels (75$\pm$49 mas).

We remind the reader that the follow-up HST observations of SN 2006my were not available to Li07 at the time of their analysis, necessitating the use of ground-based CFHT data to perform the image alignment.  The vast difference in resolution between HST and ground-based data makes it more difficult to identify common, isolated point sources that can be reliably used to align the images.  This explains the larger Li07 transformation rms and the offset between our transformed positions.  We therefore favour transformations performed using the HST follow-up data as being more accurate, both those presented in this paper and by \citet{Leonard:2008p1944}.  This is in keeping with the work of all groups involved in SN progenitor searches, including Li07, who aim to obtain the highest resolution follow-up data possible to perform differential astrometry for precisely the reasons we state above.  We stress that, had the SN 2006my HST data been available, we are confident Li07 would have obtained a similar result.

Like Li07 we used the PSF-fitting photometry package HSTphot (version 1.1.7b) \citep{Dolphin:2000p1123} to perform photometry on the pre-explosion images.  This software package includes several pre-processing tasks, which mask bad pixels, cosmic rays and hot pixels; determine the sky background and create co-added images from observations with the same pointing.  Data input to these pre-processing tasks must be the original fits files as received from the HST archive.  In order to compare our results directly with those of Li07 we were careful to choose exactly the same options as detailed in their paper.  During pre-processing we co-added the cosmic-ray split frames in each filter.  When running HSTphot we chose Option 10, which turned on local sky determination and turned off aperture corrections since there were no good aperture stars.  In such cases HSTphot applies default filter-dependent aperture corrections.  We also matched our signal-to-noise (S/N) thresholds to those of Li07 - detection S/N threshold = 2.5$\sigma$; final photometry S/N threshold = 3.0$\sigma$. 

\begin{figure}
    \centering
    \includegraphics[width=50mm]{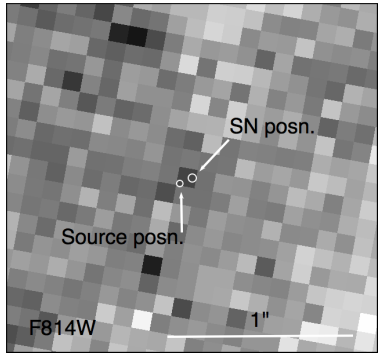}
    \caption{Magnified view of pre-explosion WFPC2 F814W image centred on the site of SN 2006my.  The transformed SN position is marked by the circle as indicated, the radius of which corresponds to the 1$\sigma$ positional uncertainty.  The position of the Li07 candidate progenitor is marked by the smaller circle, the radius of which corresponds to the positional uncertainty we derive in this paper.  This source is shown not be coincident with the SN.  The image is oriented such that North is up and East is to the left.}
\label{fig:06my_zoomin}
\end{figure}

Li07 detect a source of 5.6$\sigma$ significance in F814W with a magnitude of $m_{F814W}$=$24.47\pm0.2$, and nothing in the F555W frame.  We too find a source in F814W, with exactly the same magnitude and S/N ratio, which is indeed very close to our transformed SN position but is not coincident within the astrometric errors.  (A source is also detected in F555W but is classified by HSTphot as having a flat, extended profile and is therefore rejected from the stellar photometry list).  Having identified the source in F814W, we measured its position using the three centring algorithms\footnote{Note that the centring box used for the {\sc daophot} {\em centroiding} task was 5 pixels wide, and hence the measured source position was not erroneously shifted towards the bright pixel to the south-east in Fig.~\ref{fig:06my_zoomin}} within {\sc daophot} (see above) and used these values along with the HSTphot measurement to calculate an average pixel position: Object(x,y) = (410.40, 158.70) $\pm$0.21 WF pixels (21 mas) estimated from the range of the four measurements.  Correcting for geometric distortion \citep{Kozhurina_ISR,2003PASP..115..113A} we found Object(x$_{\rm o}$,y$_{\rm o}$) = (410.23, 158.46).

Comparing with our transformed SN position, SN(x$_{\rm o}$,y$_{\rm o}$)$_{pre}$ = (410.25, 159.21) $\pm$ 0.20 WF pixels, we find a separation of 0.75$\pm$0.29 WF pixels (75$\pm$29 mas)(see Figure~\ref{fig:06my_zoomin}).  This displacement suggests that the pre-explosion source and the SN are not coincident.

We note that the original Li07 transformation, calculated using CFHT observations of SN 2006my, did produce a SN position that was coincident with this source within the (larger) astrometric uncertainties.  Their identification of this source as a progenitor candidate was therefore perfectly valid at the time.  For the reasons stated previously, we favour the differential astrometry performed using the follow-up HST data as being more accurate.  We therefore suggest that the Li07 candidate is most probably not the progenitor of SN 2006my given the displacement we find between the SN site and the candidate source.  This agrees with the findings of \citet{Leonard:2008p1944}.  However, the ultimate test will be to re-image the explosion site after SN 2006my has faded to find out if the Li07 candidate, or indeed any other object, has disappeared (similar to the analysis of the SN 2003gd progenitor carried out by \citealt{2009Sci...324..486M}).

\subsection{Progenitor luminosity and mass limits}
\label{sec:06my_masslimits}

Since we conclude that the progenitor of SN 2006my is most probably not detected in the pre-explosion observations, we must derive detection limits for these data in order to set luminosity and mass limits on the unseen stellar pre-cursor (see, for example, \citealt{Maund:2005p492}).  The signal-to-noise ratio, S/N, of a star imaged with a CCD is given by the equation

\begin{equation}
{\rm S/N} = {\frac{F_{star}}{\sqrt{F_{star} + \sigma^2_{bg}}}}
\label{eqn:sig_noise}
\end{equation}

where $F_{star}$ is the flux of the star in units of electron counts, ${\rm e^-}$, and $\sigma_{bg}$ is the noise contribution from all other background sources other than the photon noise from the star itself ($\sqrt{F_{star}}$\,).  Here we define $\sigma_{bg}$ as

\begin{equation}
\sigma_{bg} = \sqrt{{n_{ap}}\,\left(1 + \frac{n_{ap}}{n_{sky}}\right)\,\left(F_{sky} + R^2\right)}
\label{eqn:bgnoise_value}
\end{equation}

where $n_{ap}$ and $n_{sky}$ are the number of pixels in the aperture and sky annulus respectively, $R$ is the detector readout noise in $e^-$, and ${\rm F_{sky}}$ is the mean flux of the sky background per pixel, also in $e^-$.  The term $(1 + n_{ap}/n_{sky})$ accounts for the noise incurred due to any error in the estimation of the sky background, $F_{sky}$.  This effect is mitigated by using large numbers of sky pixels to estimate the mean sky value.  Rearranging eqn.~\ref{eqn:sig_noise} and solving the resultant quadratic gives

\begin{equation}
F_{star} = \frac{{\rm(S/N)}}{2}\,\left(\,{\rm(S/N)} + \sqrt{{\rm(S/N)}^2 + 4\sigma^2_{bg}}\,\right)
\label{eqn:F_star}
\end{equation}

Taking values for all the input variables from the observations, we can use eqn. \ref{eqn:F_star} to calculate the value of $F_{star}$ required to produce a detection of a desired S/N ratio.  An apparent magnitude limit can then be calculated as

\begin{equation}
m_{lim} = -2.5\,log\,\left(\frac {F_{star}}{gain \times exptime}\right) + ZP + apcor + CTE
\label{eqn:limit_mag}
\end{equation}

where $gain$, $exptime$, $ZP$, $apcor$ and $CTE$ are the detector gain, exposure time, zeropoint, aperture correction and charge transfer efficiency correction respectively.  The updated WFPC2 Vegamag zeropoints of \citet{Dolphin:2000p1141} are taken from Andrew Dolphin's website\footnote{http://purcell.as.arizona.edu/wfpc2\_calib}.  These are defined as the magnitude of a source producing a count rate of 1 {\em Data Number} per second (1 ${\rm DN\,s^{-1}}$), which is why we covert our flux to such units in Equation~\ref{eqn:limit_mag}.  Aperture corrections are determined from the tabulated encircled energy curves of \citet{Holtzman:1995p1198} and CTE corrections \citep{Dolphin:2000p1141} calculated using the latest equations from Andrew Dolphin's website$^3$. 

Using an aperture of 2 pixels radius and a sky annulus of inner radius 10 pixels and width of 5 pixels (all WF pixels), we calculated a 3$\sigma$ detection limit for the progenitor of SN~2006my in the pre-explosion $F814W$ image: $3\sigma\,F814W$ = 24.8.  That we define a detection limit only in $F814W$ is due to the fact that we did not detect the SN progenitor.  It is obviously impossible to obtain {\em any} colour constraints from a non-detection, and hence we must assume a spectral type in order to set luminosity and mass limits.  In the case of a type II-P SN, such as SN 2006my \citep[Li07; ][]{2010MNRAS.404..981M}, we expect the progenitor to have been a red supergiant star.  SNe II-P require progenitors of large radii of the order 400-1000R$_{\odot}$ \citep{1976ApJ...207..872C,Arnett:1980p1887}, which are typical for red supergiant stars.  Of our pre-explosion observations the $F814W$ data provides by far the most restrictive limits on such a red object.

The apparent detection limit was converted to a bolometric magnitude using estimates of the distance and extinction towards the progenitor star, and bolometric and colour corrections appropriate for a red supergiant.  \citet{Solanes:2002p1225} have collected distance estimates for NGC4651 from seven different sources,  deriving a mean distance of d = 22.3 $\pm2.6$ Mpc (distance modulus $\mu = 31.74 \pm0.25$) and we adopt this value here.  Li07 found no evidence for any host-galaxy extinction in spectra of SN 2006my and hence, like them, we apply only a correction for the Galactic extinction of $E(B-V)=0.027$ assuming the reddening laws of \citet{Cardelli:1989p1230}.  Bolometric and colour corrections for an M0 supergiant were taken from \citet{2000asqu.book..381D} with a further correction between the $F814W$ and Cousins $I$ filters from \citet{Maund:2005p492}.  An uncertainty of $\pm0.3^m$ was assumed for the bolometric correction, estimated from the range of values for red supergiants of late K to late M-type \citep{Levesque:2005p1919}.  The bolometric magnitude limit was found to be ${\rm M_{bol}} = -6.13 \pm 0.39$, which corresponds to a luminosity limit of $\log L/L_{\odot} = 4.35 \pm0.16$.  Assuming the errors in the distance, extinction and bolometric correction are Gaussian in nature, one can set an 84 per cent confidence limit for the progenitor luminosity of $\log L/L_{\odot} = 4.51$.

\begin{figure}
    \centering
    \includegraphics[width=80mm]{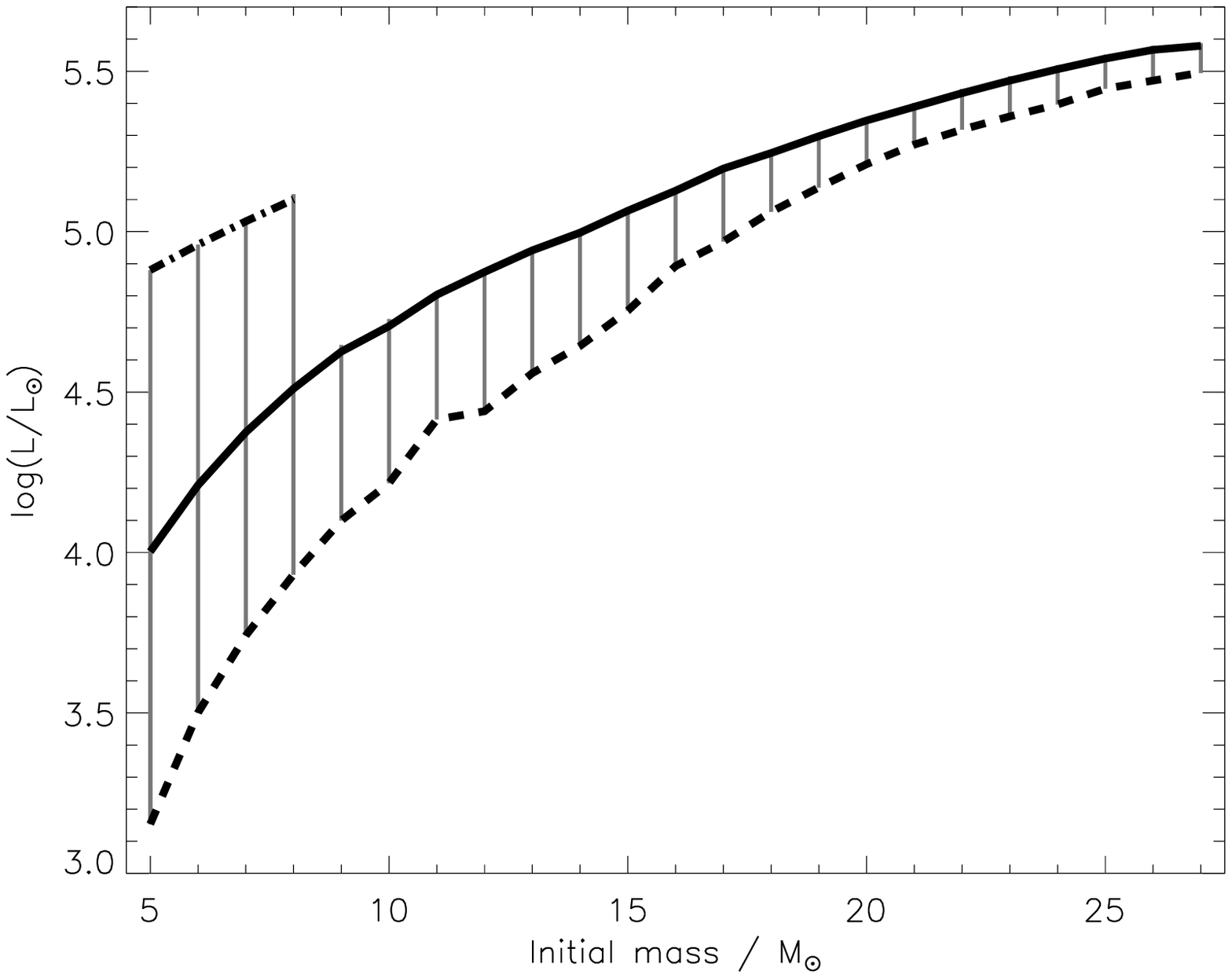}
    \includegraphics[width=80mm]{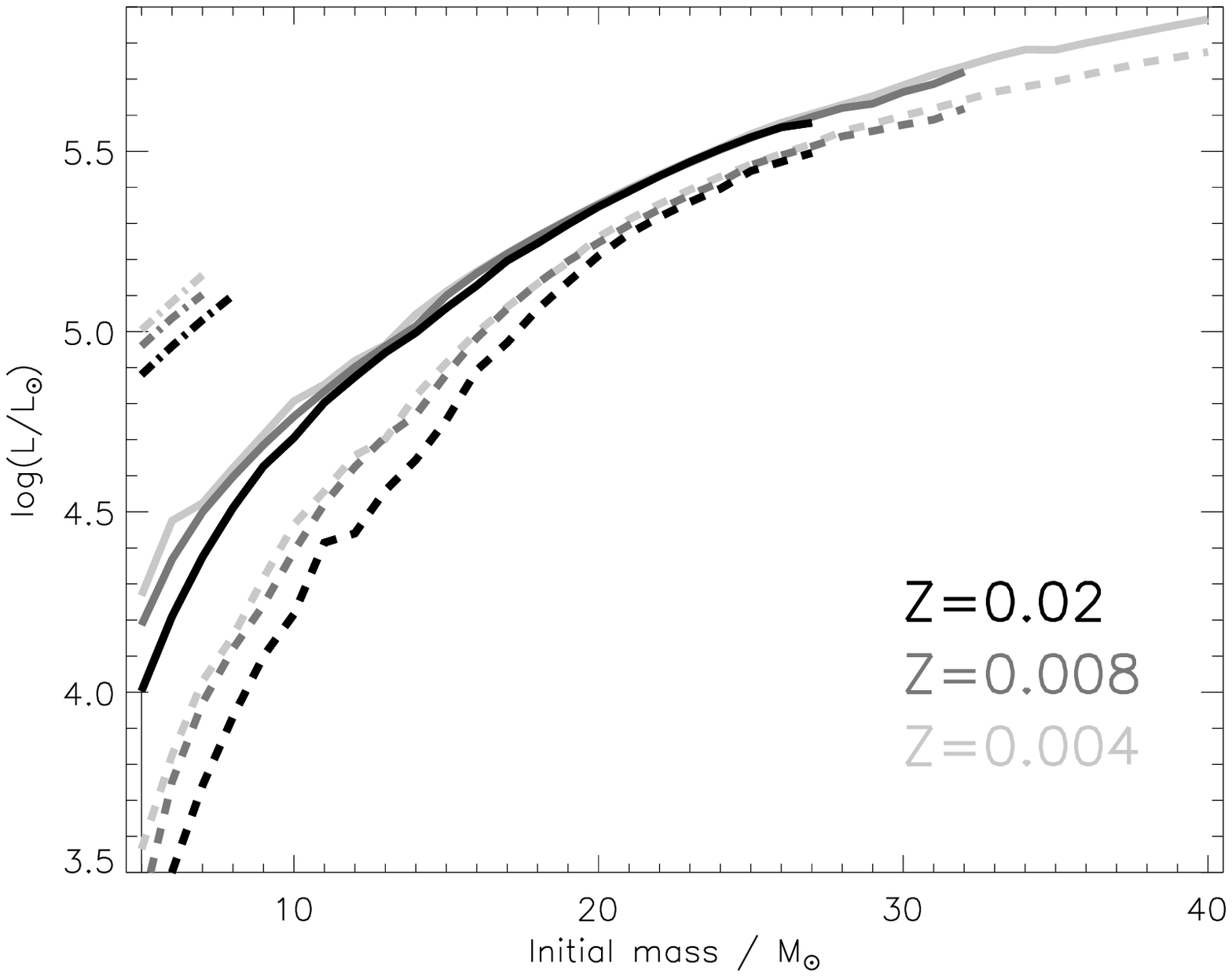}
    \caption{Reproduction of Figure 1 from \citet{2009MNRAS.395.1409S}. {\bf (a)}: Initial mass versus final luminosity of {\sc STARS} stellar models for solar metallicity ($Z=0.02$). The luminosity range for each mass corresponding to the end of He-burning and the end of the model (just before core-collapse) are marked by the thin vertical lines.  From a limit of luminosity an upper limit to the initial mass can be determined. The solid line is the luminosity of the model end-point, the dashed line the luminosity at the end of core helium burning and the dash-dotted line is the luminosity after second dredge-up when the lower mass stars become AGB stars. {\bf (b)}: The same as (a) but with three metallicities shown for comparison, and with the 
vertical joining bars omitted for clarity.}
\label{fig:STARS}
\end{figure}

Comparing this limit with the final luminosities of Cambridge {\sc stars} stellar evolutionary models \citet{Eldridge:2004p60} shown in Figure~\ref{fig:STARS} we can estimate an upper mass limit for the progenitor star.  A detailed discussion of all aspects of our methodology, including its merits over comparing progenitor photometry with the closest stellar model tracks on an HR diagram, is provided in \citet{2009MNRAS.395.1409S}.  Since metallicity affects the final luminosity of a star of given initial mass (see Figure~\ref{fig:STARS}b) we attempted to constrain the metallicity at the position of SN~2006my.  The de-projected galactocentric radius\footnote{$r_{\rm 25}$ is the radius of a galaxy at which the surface brightness drops to 25 mag per square arcsec} of SN~2006my was found to be $r_{\rm G}/r_{\rm 25}=0.37$; almost exactly the radius of the characteristic oxygen abundance measured in NGC4651 by \citet{Pilyugin:2004p1923}. They measure a value of log (O/H) + 12 = 8.7\,dex at this position, which is consistent with solar metallicity \citep[8.66$\pm$0.05]{Asplund:2004p1946}.  We note that Li07 find an oxygen abundance of $\sim$8.5\,dex and correctly interpret this as subsolar compared to the solar value of 8.8\,dex from \citet{Grevesse:1998p1924}.  With respect to the latest oxygen abundance estimates \citep[8.66$\pm$0.05]{Asplund:2004p1946} we interpret the Li07 measurement as being between our solar and LMC metallicity values.  From the LMC and solar metallicity models in Figure~\ref{fig:STARS} we derive upper initial mass limits for the progenitor of SN~2006my of 11-13M$_{\odot}$.

\section{The progenitor of SN 2006ov}
\label{sec:2006ov}

Li07 also reported the detection of a candidate red supergiant progenitor of main-sequence mass $M = 15^{+5}_{-3}$M$_{\odot}$\ for SN~2006ov in archival WFPC2 observations of M61. In this case they aligned the pre-explosion frames with HST observations of the SN.  Having performed PSF-fitting photometry using HSTphot \citep{Dolphin:2000p1123} without detecting a progenitor star, Li07 noticed that a source was still visible in the residual images close to the SN site. They conclude that this residual object is coincident with the SN position, and report that by forcing HSTphot to fit a PSF at the SN position they detect an object of 6.1$\sigma$ significance in the F814W and F450W observations.

\begin{table}
\caption{Pre-and post-explosion observations of the site of SN 2006ov}
\begin{center}
\begin{tabular}{lcrc}
\hline\hline
Date  & Telescope+Instrument & Filter  & Exp. Time \\
      &            &         & (s)       \\
\hline
\multicolumn{4}{c}{{\bf Pre-explosion images}} \\
2001 Jul 26 & HST+WFPC2 & F450W & 460\\
1994 Jun 06 & HST+WFPC2 & F606W & 160\\
2001 Jul 26 & HST+WFPC2 & F814W & 460\\
\\
\multicolumn{4}{c}{{\bf Post-explosion images}} \\
2006 Dec 12 & HST+ACS/HRC & F435W & 420\\
2006 Dec 12 & HST+ACS/HRC & F625W & 180\\
\hline\hline
\end{tabular}
\end{center}
\label{tab:06ov_obstab}
\end{table}

\subsection{Astrometry and Photometry}
\label{sec:06ov_ast_phot}

\begin{figure*}
    \centering
    \includegraphics[width=120mm]{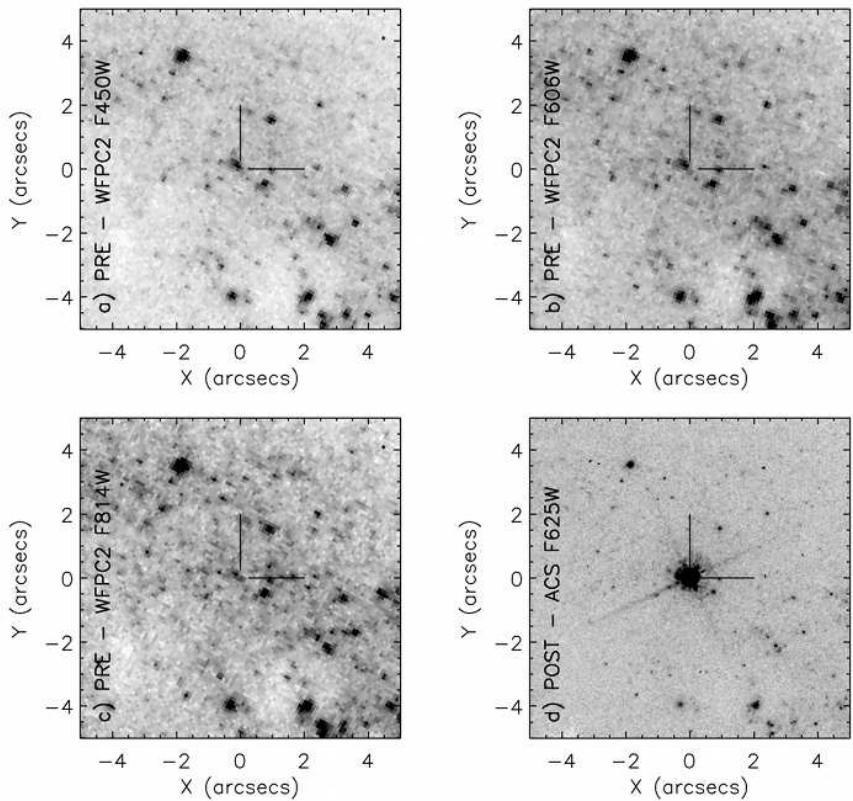}
    \caption{Aligned pre- and post-explosion HST images of the site of SN 2006ov.  All images are centred on the SN position (marked by the cross hairs) and oriented such that North is up and East is to the left.  {\bf (a)}: pre-explosion WFPC2 $F450W$, {\bf (b)}: pre-explosion WFPC2 $F606W$ {\bf (c)}: pre-explosion $F814W$ and {\bf (d)}: post-explosion ACS/HRC $F625W$ image with the SN clearly visible.  The SN position lies just south-west of an extended source in the pre-explosion frames.}
\label{fig:06ov_fig}
\end{figure*}

We have analysed the same HST data-sets as detailed in Li07 and in Table~\ref{tab:06ov_obstab} of this paper.  The pre-explosion WFPC2 data consisted of F450W and F814W images from program GO-9042 (PI: Smartt) and F606W exposures from program GO-5446 (PI: Illingworth).  The SN observations were taken with the Advanced Camera for Surveys (ACS) High Resolution Channel (HRC) (pixel scale = 0.025$\arcsec$) as part of program GO-10877 (PI: Li).  All data were downloaded from the HST archive at STScI via the OTFR pipeline, and the WFPC2 images further reduced using standard tasks within {\sc iraf} (see $\S$\ref{sec:06my_ast_phot}).  While the OTFR pipeline automatically corrected the ACS/HRC data for geometric distortion, the above reduction did not correct the WFPC2 data.  Our data reduction was therefore entirely consistent with that of Li07, except in the case of the WFPC2 F606W pre-explosion observations.  On inspection of this data-set we noticed that the pointings of the cosmic-ray (CR) split exposures were offset.  With respect to the coordinate frame of the WF4 chip (on which the SN site was imaged), these offsets (exposure 1 {\em minus} exposure 2) were measured to be $\delta$x = -0.12 pix and $\delta$y = 1.16 pix.  The CR-split frames were first aligned, using exposure 1 as the reference frame, and then combined to produce the final reduced image (see Figure~\ref{fig:06ov_fig}).  The ``clipped'' nature of objects along the y-axis of the co-added F606W image in Li07 (see Fig. 8 in Li07), suggests that these offsets were missed.  Li07 also used the HSTphot task {\em coadd} to combine the CR-split exposures before running the photometry package.  This task cannot apply image offsets and will therefore have produced a similarly clipped image.  Finally, the coordinate transformation we derive for this pre-explosion image results in a progenitor position that is offset in the y-direction from that of Li07 (see below).  This is easily explained as a consequence of the differences in our respective reductions of the $F606W$ data.

We have repeated the alignment of the pre- and post-explosion HST observations originally performed by Li07, except that here, as in $\S$\ref{sec:06my_ast_phot}, we have applied the geometric distortion corrections of \citet{2003PASP..115..113A} and \citet{Kozhurina_ISR} to the WFPC2 coordinates prior to calculating the transformation.   

In all of the pre-explosion observations the SN site was imaged on the WF4 chip of the WFPC2 mosaic.  The $F450W$ and $F814W$ WFPC2 observations were found to be well registered and hence a single transformation was deemed appropriate for both.  The positions of 31 sources common to both the pre-explosion $F814W$ and the post-explosion $F625W$ images were measured, corrected for distortion, and subsequently used to calculate a transformation (x and y shifts, scales and rotations) between the pre- and post-coordinate frames.  This transformation had an rms error of $\pm$0.15 WF pixels ($\pm$15 mas).  A separate transformation between the pre-explosion $F606W$ image\footnote{We assume a distortion correction appropriate for the $F555W$ image \citep{2003PASP..115..113A}} and the post-explosion frame was calculated from the positions of 31 common objects, with a total rms error of $\pm$0.15 WF pixels ($\pm$15 mas).  

We tested the robustness of the transformation rms values using the same method as employed in $\S$\ref{sec:06my_ast_phot}.  The results of these tests are shown in Table~\ref{tab:06ov_rms}.  The residuals calculated from the fit-independent coordinates were only slightly larger than the {\em geomap} residuals, suggesting that there was negligible over-fitting by the {\em geomap} task, and that distortion corrections had been successful.  Nevertheless, we adopted the slightly larger $\pm$0.17 WF pixels ($\pm$17 mas) and $\pm$0.16 WF pixels ($\pm$16 mas) as conservative, total rms errors for the WFPC2 F814W and WFPC2 F606W transformations respectively.

\begin{table}
\caption{Fit residuals for {\sc iraf} {\em geomap} transformations between pre- and post-SN 2006ov HST observations.}
\begin{center}
\begin{tabular}{rcrr}
\hline\hline
\multicolumn{4}{l}{ACS/HRC F625W to WFPC2 F814W}\\
31 source transform &\vline &16 source tranform & 15 source transform\\
$\sigma_x$ = 0.081 &\vline &$\sigma_x$ = 0.077 & $\sigma_x$ = 0.085 \\
$\sigma_y$ = 0.122 &\vline &$\sigma_y$ = 0.125 & $\sigma_y$ = 0.106 \\
$\sigma_{total}$ = 0.146 &\vline &$\sigma_{total}$ = 0.147 & $\sigma_{total}$ = 0.136 \vspace{3mm}\\
&\vline&\multicolumn{2}{l}{(residuals estimated from transformed positions}\\
&\vline&\multicolumn{2}{l}{of fit-independent source coordinates)}\\
&\vline &$\sigma_x$ = 0.087 & $\sigma_x$ = 0.079 \\
&\vline &$\sigma_y$ = 0.137 & $\sigma_y$ = 0.145 \\
&\vline &$\sigma_{total}$ = 0.162 & $\sigma_{total}$ = 0.165\vspace{5mm}\\
\hline
\multicolumn{4}{l}{ACS/HRC F625W to WFPC2 F606W}\\
31 source transform &\vline &16 source tranform & 15 source transform\\
$\sigma_x$ = 0.095 &\vline &$\sigma_x$ = 0.076 & $\sigma_x$ = 0.107 \\
$\sigma_y$ = 0.114 &\vline &$\sigma_y$ = 0.125 & $\sigma_y$ = 0.100 \\
$\sigma_{total}$ = 0.148 &\vline &$\sigma_{total}$ = 0.146 & $\sigma_{total}$ = 0.146 \vspace{3mm}\\
&\vline&\multicolumn{2}{l}{(residuals estimated from transformed positions}\\
&\vline&\multicolumn{2}{l}{of fit-independent source coordinates)}\\
&\vline &$\sigma_x$ = 0.122 & $\sigma_x$ = 0.091 \\
&\vline &$\sigma_y$ = 0.107 & $\sigma_y$ = 0.129 \\
&\vline &$\sigma_{total}$ = 0.162 & $\sigma_{total}$ = 0.158\\
\hline\hline
\multicolumn{3}{l}{All values are in units of WF chip pixels (0.1$\arcsec$ / pix)}&\\
\end{tabular}
\end{center}
\label{tab:06ov_rms}
\end{table}

The position of SN~2006ov in the post-explosion ACS/HRC F625W image was measured using the three centring algorithms within {\sc daophot} and the PSF-fitting photometry package DOLPHOT.\footnote{http://purcell.as.arizona.edu/dolphot/}$^,$\footnote{Positions measured by DOLPHOT differ from those of {\sc daophot} by -0.5 in both x and y coordinates.  All positions quoted in this paper are corrected to {\sc daophot} values.}  The mean of the four measurements was adopted as the SN position, SN(x$_{\rm o}$,y$_{\rm o}$)$_{post}$ = (598.61, 613.89) $\pm$0.12 ACS/HRC pixels (3 mas), estimated from the range of the four measurements.  We again note that the ACS image was corrected for geometric distortion during pipeline reduction, hence no further correction was required.

\begin{table}
\caption{Comparison of transformed SN 2006ov positions in pre-explosion WFPC2 data from Li07 and this paper.  Li07 coordinates corrected for distortion by applying the corrections of \citet{2003PASP..115..113A} and \citet{Kozhurina_ISR}.}
\begin{center}
\begin{tabular}{lccc}
\hline\hline
Pre-explosion& Li07 position & Li07 position & Crockett et al. position\\
image& & (distortion corrected) & (distortion corrected) \\
\hline
F450W/F814W & (571.62, 236.22) $\pm$0.17 & (571.76, 236.09) $\pm$0.17 & (571.69, 236.19) $\pm$0.17 \vspace{3mm} \\
F606W & (227.44, 267.34) $\pm$0.16 & (227.64, 267.44) $\pm$0.16 & (227.81, 268.01) $\pm$0.16 \\
\hline\hline
\multicolumn{4}{l}{Coordinates are for WF4 chip (0.1$\arcsec$ / pix), while uncertainties are total values (x and y errors combined)}\\
\end{tabular}
\end{center}
\label{tab:06ov_SNposn}
\end{table}

Applying the 31 source ACS-to-WFPC2 F814W transformation, we found a distortion corrected position for SN~2006ov on the pre-explosion $F450W$ and $F814W$ images of SN(x$_{\rm o}$,y$_{\rm o}$)$_{pre}^{F450W/F814W}$ = (571.69, 236.19) $\pm$0.17 WF pixels (17 mas).  This can be considered coincident with the transformed SN position from Li07, being offset by just 0.12$\pm$0.24 WF pixels after correcting the Li07 coordinates for geometric distortion (see Table~\ref{tab:06ov_SNposn}). 

However, this is not the case for the $F606W$ pre-explosion image.  Applying the 31 source ACS-to-WFPC2 F606W transformation, we found the transformed distortion corrected SN position to be SN(x$_{\rm o}$,y$_{\rm o}$)$_{pre}^{F606W}$ = (227.81, 268.01) $\pm$0.16 WF pixels (16 mas), which is offset by 0.59$\pm0.23$ WF pixels from the distortion corrected Li07 position (see Table~\ref{tab:06ov_SNposn}).  This difference is due to the shift between the F606W CR-split images that was missed by Li07 during data reduction.  We therefore adopt our F606W transformation as the most reliable.

HSTphot was used to perform PSF-fitting photometry on the WFPC2 images.  HSTphot and all pre-processing steps were run in precisely the same manner as in Li07 (see $\S$\ref{sec:06my_ast_phot}), except in the case of the $F606W$ data where the offset between the CR-split exposures required special treatment.  Here the exposures were not co-added; rather they were input separately to HSTphot along with appropriate offsets, using the first exposure as the reference frame.  HSTphot returned no objects coincident with the SN position in any of the pre-explosion frames, but a point-like source was discovered {\em close} to the SN site in the $F814W$ residual image.  Given that we have used identical HSTphot settings, we assume that this is the same $F814W$ residual as observed by Li07, which they subsequently identified as a candidate progenitor for SN 2006ov.  We, however, cannot conclude that this source is {\em definitely} coincident with the SN position.

\begin{figure*}
    \centering
    \begin{tabular}{ccc}
    \includegraphics[width=50mm]{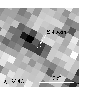} &
    \includegraphics[width=50mm]{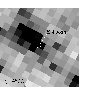} &
    \includegraphics[width=50mm]{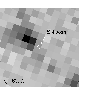} \\
    \includegraphics[width=50mm]{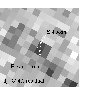} &
    \includegraphics[width=50mm]{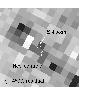} &
    \includegraphics[width=50mm]{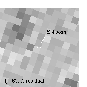}                
    \end{tabular}
    \caption{{\bf (a-c)}: Zoom-in of SN 2006ov site on pre-explosion WFPC2 $F814W$, $F450W$ and $F606W$ images.  The transformed position of the SN is marked with a circle, the radius of which represents the 1$\sigma$ positional uncertainty.  {\bf (d-f)} Corresponding residual images, produced by HSTphot after subtraction of all detected sources.  All images are oriented such that North is up and East is to the left.  A point-like source is seen close to the SN site in the $F814W$ residual frame (d).  Its position, marked by the smaller circle, is not coincident with the SN within the combined uncertainties.  An extended and less significant residual source is seen in the $F450W$ frame (e).  Again its centre is not coincident with the transformed SN position.  No residual source is visible at, or near, the SN position in F606W (f).  See text for details.}
\label{fig:res_zoom}
\end{figure*}

Figure~\ref{fig:res_zoom} shows magnified sections of the WFPC2 pre-explosion frames along with the corresponding HSTphot residual images.  All are centred on the transformed SN position, marked by a circle which represents the 1$\sigma$ astrometric uncertainty.  

The point-like $F814W$ residual can be seen close to the centre of Figure~\ref{fig:res_zoom}d.  We measured its position to be Res(x,y)$_{F814W}$ = (571.16, 235.80) $\pm$0.12 WF pixels, which is marked in Figure~\ref{fig:res_zoom}d by the smaller of the two circles just south of the SN position.  Note that this is the {\em raw} position of the residual source, not corrected for geometric distortion.  These coordinates can therefore be compared directly with those quoted in Li07.  Comparing our residual position to the Li07 transformed SN position, we found that the two are separated by 0.62$\pm$0.21 WF pixels.  Correcting the residual position for distortion, Res(x$_{\rm o}$,y$_{\rm o}$)$_{F814W}$ = (571.30, 235.67), and comparing with our own transformed SN position (Table~\ref{tab:06ov_SNposn}) we found the same offset of 0.62$\pm$0.21 WF pixels.    

The $F450W$ residual source (Figure~\ref{fig:res_zoom}e) is neither point-like, nor as significant as that in $F814W$.  If it is a real source, its extended nature would suggest that it is not a single object.  Nonetheless we have attempted to determine its centroid, Res(x,y)$_{F450W}$ = (571.17, 235.55) $\pm$0.13 WF pixels, which is marked on Figure~\ref{fig:res_zoom}e and lies 0.81$\pm$0.21 pixels from the Li07 transformed SN position.  Again, correcting the residual position for geometric distortion, Res(x$_{\rm o}$,y$_{\rm o}$)$_{F450W}$ = (570.86, 235.91), and comparing with our transformed SN position we found an offset of 0.88 $\pm$0.21 WF pixels.

The shallower $F606W$ observation (Figures~\ref{fig:res_zoom}c\&f) shows no apparent residual at, or near, the SN site.

Li07 report that by forcing HSTphot to fit a PSF at their transformed SN position they detect a source of 6.1$\sigma$ significance in $F450W$ and $F814W$, and 2.2$\sigma$ in F606W.  We performed the same forced-fitting, being careful to convert all {\sc daophot} positions to HSTphot coordinates by subtracting 0.5 pixels from the x and y values.  We were unable to reproduce their photometry at the transformed SN site, finding detections of 3.6$\sigma$ and 2.9$\sigma$ significance in $F814W$ and $F450W$ respectively, and nothing in $F606W$.  

Detections of the highest significance ($F450W$ = 24.16$\pm$0.24 (4.5$\sigma$);  $F814W$ = 23.36$\pm$0.18 (6.0$\sigma$)) were found when forcing PSF fitting at our Residual source positions, which are offset from the transformed SN position.  Admittedly, it is difficult to judge by eye which of the two circles, the SN position or the Residual centroid, is closer to the {\em true} centre of the residual source in the panels of Figure~\ref{fig:res_zoom}.  However, since our best-fit photometry arises when fitting at the measured Residual coordinates, we favour this as its true position.

Since we cannot reproduce the photometry of Li07 at the transformed SN site, and have found that the residual source is offset from the SN position, we suggest that this object is probably not the progenitor star.  Given its proximity to the SN site, it is likely that some of the pre-explosion source flux in contributed by the progenitor.  However, the evidence shown here suggests that it would be wrong to attribute all of this flux to the progenitor.  As the residual itself is close to the noise limit in both $F450W$ and $F814W$, we suggest that the progenitor of SN 2006ov is not observed.  The ultimate test will be to re-image the explosion site with HST after SN 2006ov has faded to find out if the Li07 candidate, or indeed any other object, has disappeared.

\subsection{Progenitor luminosity and mass limits}

Luminosity and mass limits were set for the assumed red supergiant progenitor following the method described in $\S$\ref{sec:06my_masslimits}.  The sky background at the SN site was estimated and equations~\ref{eqn:bgnoise_value}, \ref{eqn:F_star} \& \ref{eqn:limit_mag} used to calculate the 3$\sigma$ limiting magnitude of $3\sigma\,F814W$ = 24.2.

Estimates of the distance, extinction and bolometric and colour corrections were applied to convert this detection limit to a bolometric magnitude.  Li07 calculated a mean distance modulus for M61 of $\mu = 30.5 \pm0.4$ ($d = 12.6 \pm2.4$ Mpc) from two Tully-Fisher distance estimates \citep{1988ngc..book.....T,Schoeniger:1997p1926}.  They also found no evidence for any host-galaxy extinction in spectra of SN 2006ov, applying only a small correction of $E(B-V)=0.022$ for Galactic extinction.  We adopt both these values here, and make the same assumptions for the bolometric and colour corrections as we have done in $\S$\ref{sec:06my_masslimits}.  In this way we derived a $3\sigma$ bolometric magnitude limit of $M_{bol} = -5.48 \pm 0.50$, corresponding to a luminosity limit of $\log L/L_{\odot} = 4.09 \pm0.20$.  Assuming the errors are Gaussian, an upper luminosity limit of 84 per cent confidence can be set of $\log L/L_{\odot} = 4.29$.

The oxygen abundance gradient of M61 was re-defined by \citet{Pilyugin:2004p1923} and we use this, along with the galactocentric radius $r_{\rm G}/r_{\rm 25}=0.26$, to estimate an oxygen abundance of 8.9\,dex at the position of SN~2006ov.  Hence, we compare the above luminosity limit with {\sc stars} models of solar metallicity in Figure~\ref{fig:STARS}, finding an upper initial mass limit for the SN progenitor of 10M$_{\odot}$.

\section{The progenitor of SN 2004et}
\label{sec:2004et}

A candidate progenitor star for the type II-P SN 2004et was identified by Li05 on archival images from the Canada France Hawaii Telescope (CFHT). These included $BVR$ CFH12K images from 2002 and u'g'r' MegaCam observations from 2003. The pre-explosion images were aligned with SN observations from the Katzman Automatic Imaging Telescope (KAIT), yielding a transformation between the pre- and post-explosion coordinate systems with a rms error of $\sim 0.1\arcsec$. The candidate progenitor was found to be coincident with the SN position to within these astrometric errors. Through comparision with stellar evolutionary models Li05 characterised this source as a single, {\em yellow} supergiant star with an initial main sequence mass of 15$^{+5}_{-2}$M$_{\odot}$. Both the mass and the colour of this candidate progenitor were of note; the mass since it was the largest detected for a normal II-P SN, and the colour since such progenitors were expected to be {\em red} not {\em yellow} supergiants. 

To more accurately determine the position of SN 2004et on the pre-explosion frames \citet{Li:2005p1927} aligned the CFHT data with HST ACS/HRC observations of the SN.  They reported that the transformed position of the SN was ``consistent with the progenitor in the CFHT image to within $0.056\arcsec \pm0.043\arcsec$'' and that this confirmed their identification of the progenitor star.  

We have since obtained deep follow-up observations of the site of SN 2004et from the William Herschel Telescope (WHT), the HST and the Gemini (North) observatory, which show that the Li05 candidate progenitor is still visible approximately 3 years post-explosion, and that it is in fact resolved into at least 3 separate sources.  We also present hitherto unpublished pre-explosion $i'$-band observations of the SN site, taken at the Isaac Newton Telescope (INT), in which we detect an excess flux that we attribute to the unresolved progenitor of SN 2004et.

\subsection{Data acquisition and reduction}
\label{sec:astrometry}

\begin{table}
\caption{Pre-and post-explosion observations of the site of SN 2004et.} 
\begin{center}
\begin{tabular}{lcrc}
\hline\hline
Date  & Telescope+Instrument & Filter  & Exp. Time \\
      &            &         & (s)       \\
\hline
\multicolumn{4}{l}{{\bf Pre-explosion images}} \\
2002 Aug 06 & CFHT+CFH12K & B & 450\\
2002 Aug 06 & CFHT+CFH12K & V & 300 \\
2000 Sep 27 & CFHT+CFH12K & R & 360 \\
2002 Aug 12 & INT+WFC & i' & 3600 \\
\\
\multicolumn{4}{l}{{\bf Post-explosion images showing bright SN}} \\
2005 May 02 & HST+ACS/HRC & F625W & 360 \\
\\
\multicolumn{4}{l}{{\bf Late-time post-explosion images}} \\
2007 Aug 12 & WHT+AUX & B & 2400 \\
2007 Aug 12 & WHT+AUX & V & 1800 \\
2007 Aug 12 & WHT+AUX & R & 1800 \\
2007 Aug 12 & WHT+AUX & I & 2400 \\
2007 Jul 08 & HST+WFPC2 & F606W & 1600 \\
2007 Jul 08 & HST+WFPC2 & F814W & 1600 \\
2007 Jul 08 & HST+NICMOS & F110W & 640 \\
2007 Jul 08 & HST+NICMOS & F160W & 512 \\
2007 Jul 08 & HST+NICMOS & F205W & 576 \\
2008 Jan 19 & HST+WFPC2 & F606W & 1600 \\
2008 Jan 19 & HST+WFPC2 & F814W & 1600 \\
2008 Jan 19 & HST+NCMOS & F110W & 640 \\
2008 Jan 19 & HST+NICMOS & F160W & 512 \\
2008 Jan 19 & HST+NICMOS & F205W & 576 \\
2008 Jul 10 & Gemini-N+Altair/NIRI & K & 7200 \\
\hline\hline
\end{tabular}
\end{center}
\label{tab:04et_obstab}
\end{table}

The pre- and post-explosion observations for SN 2004et are detailed in Table~\ref{tab:04et_obstab}.  The Canada France Hawaii Telescope (CFHT) datasets were downloaded from the CFHT archive at the Canadian Astronomy Data Centre (CADC)\footnote{http://www4.cadc-ccda.hia-iha.nrc-cnrc.gc.ca/cadc} along with appropriate flat field and bias frames.  The $B$\&$V$ data were the same as those analysed by Li05.  However, the $R$-band image was different.  Taken in 2000, the total exposure time of 360 sec is only slightly longer than that of the 2002 $R$-band image (300 sec) analysed by Li05, but the image quality is $0.6\arcsec$ (compared to $0.8\arcsec$) and the SN site happened to fall on one of the high resistivity (HiRho) bulk silicon chips of the CFH12K mosaic.  The sensitivity of these chips in the $R$-band is on average 15 percent higher than the standard epitaxial silicon (EPI) chips that make up the rest of the mosaic, and on which the observations from Li05 were taken.  The net result is a deeper pre-explosion image than has yet been published.  All CFHT data were reduced using standard tasks within {\sc iraf}.  In addition we discovered a deep INT+WFC i'-band observation from 2002, not analysed by Li05, which extended the pre-explosion SED to longer wavelengths.  The INT data were downloaded from the CASU INT Wide Field Survey (WFS) \footnote{http://www.ast.cam.ac.uk/~wfcsur}, having been processed and calibrated via the INT WFC pipeline.  The HST Advanced Camera for Surveys/High Resolution Channel (ACS/HRC) observation of SN 2004et from 2005 was retrieved from the HST archive at STScI via the OTFR pipeline.  The pipeline Multidrizzle task combined the sub-exposures and corrected the resultant image for geometric distortion.

We re-observed the field of SN 2004et using the Auxilary Port Imager (AUX) on the WHT in Aug 2007, $\sim$3 yrs post-explosion.  By this late-epoch we judged the SN would have faded sufficiently to allow us to check whether or not the Li05 candidate progenitor had disappeared; a key test of its validity.  The observations were taken in $BVRI$ filters to match as closely as possible the pre-explosion CFHT and INT images.  All WHT data were reduced using standard tasks within {\sc iraf}.

In an attempt to resolve the explosion site at late-time, after the SN had faded, ground-based adaptive optics (AO) observations of the field of SN~2004et were taken on 2008 July 10 ($\sim$ 4 yrs post-explosion) using Altair/NIRI on the 8.1-m Gemini (North) Telescope (programs GN-2008A-Q-28, PI: Crockett and GN-2008A-Q-26, PI:Gal-Yam).  It was only afterwards that the authors discovered proprietary HST WFPC2 and NICMOS data, taken 2007 July 8, had become public 2 days earlier.  Further HST data, taken 2008 Jan 19, became public in Jan 2009.  

The Gemini observations were taken using a nearby 2MASS source (2MASS 20352132+6007102) as the natural guide star (NGS) for the Altair AO wave front sensor.  This guide star was separated by around 30$\arcsec$ from the SN position.  Due to the geometry of the detector and wavefront sensors, the maximum displacement of a guide star from the centre of the FOV is 25$\arcsec$.  The site of SN~2004et was therefore imaged off-centre using the f/14 camera, which has a FOV of 51$\arcsec\times51\arcsec$ (pixel scale of 0.05$\arcsec$).  The effects of anisoplanatism, which would seriously reduce the image quality at such large separations from the guide star, were mitigated by inclusion of the field lens in the optical path.  A series of short exposures totalling 7200 sec on-source were taken in the $K$-band and subsequently reduced, sky subtracted and combined using the NIRI reduction tools within the {\sc iraf gemini} package.  The quality of the reduced image at our target position was 0.15$\arcsec$; excellent given the large displacement from the AO guide star and that the guide star itself was relatively faint for its purpose.

The HST WFPC2 and NICMOS observations were carried out as part of program GO-11229 (PI: Meixner).  All data were downloaded from the STScI archive via the OTFR pipeline.  WFPC2 observations were taken in F606W and F814W filters, in each case as a series of 4$\times$400 sec exposures.  NICMOS F110W, F160W and F205W data were taken as 5$\times$128 sec, 4$\times$128 sec and 4$\times$144 sec exposures respectively.  Both the WFPC2 and NICMOS sub-exposures were dithered, and we used the {\em drizzle} technique \citep{Fruchter:2002p1494} to combine the sub-exposures in each filter.  This reduction process also corrected the WFPC2 and NICMOS images for geometric distortion.  The SN site was imaged on the PC chip of the WFPC2 instrument (drizzled pixel scale = 0.0455$\arcsec$) and on the NIC2 camera of NICMOS (drizzled pixel scale = 0.075$\arcsec$). 

In the following subsections we discuss first our natural-seeing, ground-based observations, following by the HST and Gemini images, before drawing final conclusions from the combined data set.

\subsection{Ground-based, seeing-limited data: image alignment and photometry}
\label{sec:ground_phot}

\begin{figure*}
    \centering
    \includegraphics[width=120mm]{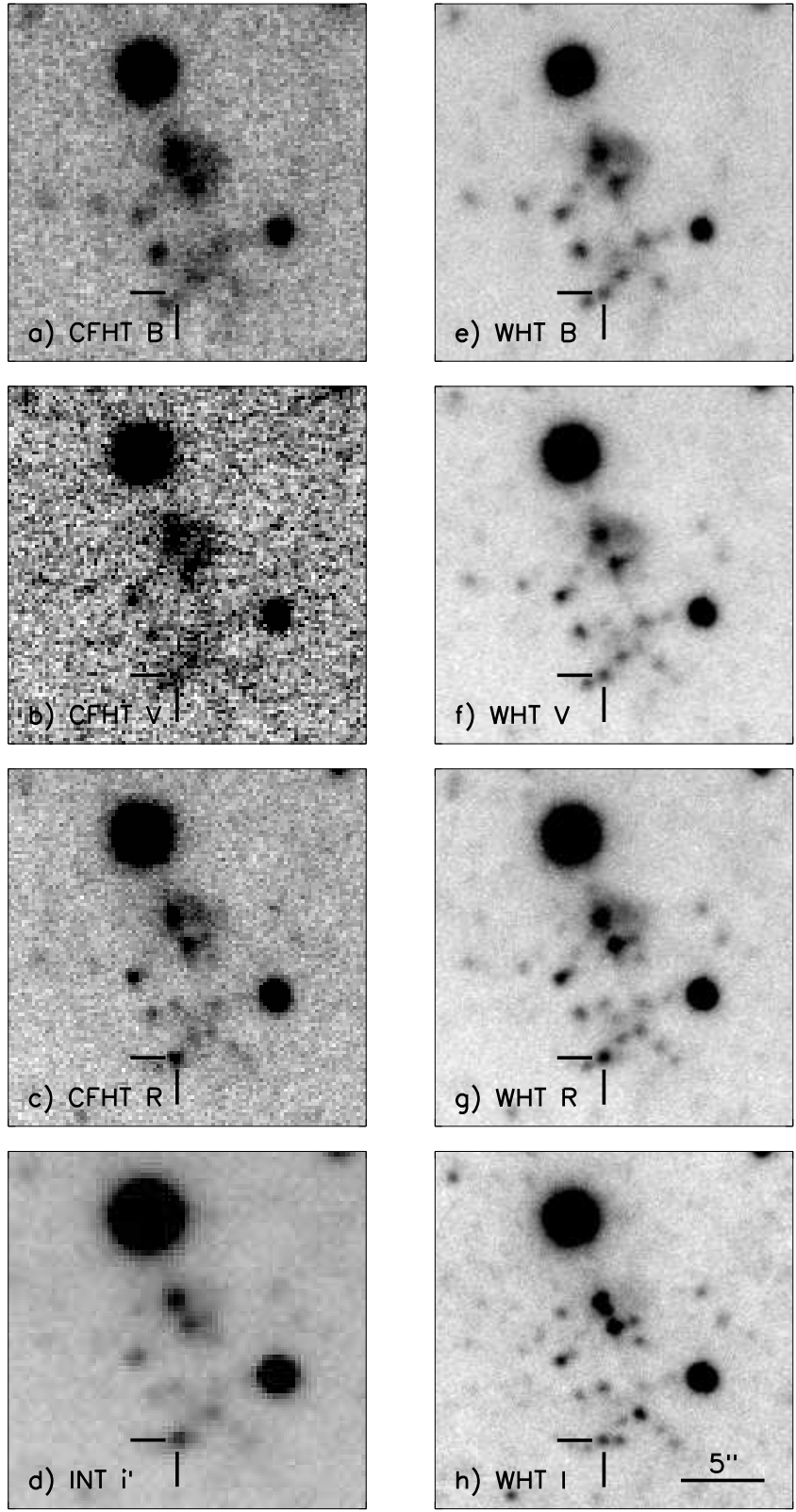}
    \caption{Archival pre-explosion and late-time post-explosion observations of the field of SN 2004et.  {\bf (a\&b)}: pre-explosion CFHT+CFH12K $B$\&$V$ observations, as studied previously by Li05; {\bf (c)}: pre-explosion CFHT+CFH12K $R$ image of higher image quality and depth than previously studied; {\bf (d)}: pre-explosion INT+WFC sloan-i' image {\bf (e-h)}: WHT+AUX late-time $BVRI$ observations taken $\sim$ 3 years post-explosion when the SN has faded.  The Li05 progenitor candidate is marked by the cross hairs.  Note that this object is still visible post-explosion in $BVR$ with similar magnitudes (see Table~\ref{tab:prepost_phot}), proving that it is not {\em solely} the SN progenitor at these wavelengths.  However, in the $I$-band the object is significantly brighter pre-than post-explosion.  This excess is due to the unresolved progenitor star.  All images are aligned and oriented such that North is up and East is to the left.}
\label{fig:pre-post}
\end{figure*}

\begin{figure}
    \centering
    \includegraphics[width=60mm]{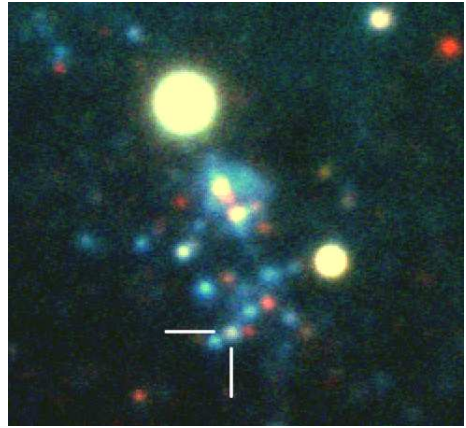}
    \caption{Pseudo-colour image created from WHT $BVI$ observations of the site of SN 2004et, taken $\sim$3 yrs post-explosion.  The cross hairs mark the position of the object originally identified by Li05 as the candidate SN progenitor.  This yellow object is still visible 3 years after the explosion.  Oriented such that  North is up and East is to the left.}
\label{fig:colour_post}
\end{figure}

In Figure~\ref{fig:pre-post} we show the CFHT+CFH12K $B\&V$ images, as presented by Li05; the previously unpublished CFHT+CFH12K $R$ observation from 2000; and the deep INT+WFC i'-band observation from 2002, also previously unpublished.  In the same figure we present our late time WHT+AUX $BVRI$ observations, with the $BVI$ images combined in Figure~\ref{fig:colour_post} to create a pseudo-colour image.  It is immediately apparent that an object is still visible at the position of the Li05 candidate progenitor in all bands.  Alignment of the CFHT and WHT $R$-band images demonstrated that the pre- and post-explosion sources are coincident, having a measured separation of $5 \pm 50$ mas. 

%\citet{2009ApJ...704..306K} found that, starting October 2007, SN~2004et {\em did} in fact re-brighten significantly in mid-IR Spitzer data.  The HST and Gemini data from January and July 2008, and presented later in this paper, show a similar re-brightening in the near-IR ($\S$\ref{sec:echo}).  However, previous mid-IR observations from August 2007 (around the time of our WHT observations) had still shown a declining light curve \citep{2009ApJ...704..306K}, so our assumption above may not be unreasonable.  We will discuss the IR-echo in more detail in $\S$\ref{sec:echo}.

\begin{figure}
    \centering
    \includegraphics[width=70mm]{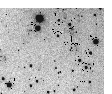}
    \caption{Pre-explosion CFHT+CFH12K $B$ image of the site of SN 2004et.  Field standard stars with calibrated magnitudes taken from \citet{Misra:2007p1117} and Li05 are labelled (see Table~\ref{tab:04et_phot_standards}).  North is up and East is to the left.}
\label{fig:04et_phot_standards}
\end{figure}

PSF-fitting photometry of the Li05 candidate progenitor source was carried out on both the archival and late-time observations using tasks within the {\sc iraf daophot} package.  Although photometry for the CFH12K $B \& V$ images was presented by Li05, it was decided to repeat this analysis so as to obtain measurements in a consistent fashion across all our pre- and post-SN observations.  Zeropoints and colour corrections were derived using stars in the field with calibrated magnitudes taken from Li05 and \citet{Misra:2007p1117} (see Figure~\ref{fig:04et_phot_standards} and Table~\ref{tab:04et_phot_standards}).  The INT+WFC i'-band observation was an exception, the zeropoint being taken from the image header (calculated during the WFS calibration process) and colour transformations to the Landolt system from the INT Wide Field Survey (WFS) website\footnote{http://www.ast.cam.ac.uk/~wfcsur/technical/photom/colours/}.  A series of Landolt standard stars were observed every night during the INT WFS, and secondary standards were also established in each field.  The photometric calibration of the WFS data is therefore very reliable.  Nevertheless we have checked this calibration using the secondary standards from Li05 and \citet{Misra:2007p1117} presented in Table~\ref{tab:04et_phot_standards}, and found that the results are consistent.  Several of the isolated standard stars (Figure~\ref{fig:04et_phot_standards}) were used to build emperical PSFs for each image.  Sources close to Li05 candidate progenitor were fitted simultaneously, but their photometry is not detailed in this paper.  Our pre- and post-explosion photometry is recorded in Table~\ref{tab:prepost_phot}, along with the pre-explosion photometry of Li05. 

\begin{table}
\caption{Calibrated photometry of secondary standards in the field of SN 2004et.  Photometry taken from \citet{Misra:2007p1117} and Li05. ID numbers correspond to those on Figure~\ref{fig:04et_phot_standards}. Photometric errors are in parentheses.}
\begin{center}
\begin{tabular}{lcccc}
\hline\hline
ID  & $B$ & $V$  & $R$ & $I$\\
\hline
1 & 15.80(0.008) & 14.73(0.010) & 14.14(0.005) & 13.47(0.023)\\
2 & 15.51(0.008) & 14.15(0.009) & 13.41(0.004) & 12.63(0.023)\\ 
3 & 19.40(0.027) & 18.48(0.015) & 17.96(0.013) & 17.31(0.053)\\
4 & 19.44(0.040) & 18.29(0.030) & 17.60(0.020) & -\\
5 & 18.31(0.017) & 17.10(0.012) & 16.43(0.018) & 15.69(0.026)\\
6 & 15.56(0.008) & 14.28(0.009) & 13.63(0.004) & 12.81(0.023)\\ 
7 & 18.30(0.018) & 17.46(0.013) & 17.04(0.005) & 16.39(0.030)\\
8 & 18.43(0.018) & 17.66(0.013) & 17.28(0.043) & 16.67(0.040)\\
9 & 18.33(0.015) & 17.39(0.011) & 16.93(0.007) & 16.25(0.028)\\
\hline\hline
\end{tabular}
\end{center}
\label{tab:04et_phot_standards}
\end{table}

\subsubsection{V \& R-band photometry}
\label{sec:VR_phot}

Comparing the pre-explosion photometry of the Li05 candidate progenitor with that of the coincident source at late-time, we find that measurements in the $V$ \& $R$ bands match within the errors.  

There are three possible solutions to these observations: 1) that the $V$ \& $R$ flux of the pre- and post-explosion objects come from the same source or blended sources, with negligible contributions from the progenitor and SN respectively; 2) the pre-SN $V$ \& $R$ flux is entirely from the progenitor star while the late-time flux is from the SN; 3) the object consists of multiple blended sources with the progenitor and SN contributing some significant fraction of the pre- and post-explosion $V$ \& $R$ flux respectively.

The second scenario seems the least likely as it requires the SN to be significantly brighter than expected (predicted $V\approx$26.1, and $R\approx$26.2 by extrapolating the late-time light curves of \citealt{2010MNRAS.404..981M}) while at the same time coincidentally matching the $V \& R$ magnitudes of its progenitor, and is ultimately ruled out by the HST and Gemini AO which resolves the object at the SN site into 3 distinct objects ($\S$\ref{sec:high_res_04et}).  The third scenario still requires a certain amount of {\em fine-tuning} in order to match the pre- and post-SN photometry.  However, the degree of fine-tuning can be reduced if we arbitrarily assume ever increasing flux contributions from the other constituent sources, so that it becomes possible to ``wash-out'' differences between the progenitor and SN magnitudes.  The first scenario is the simplest, and assumes that the progenitor and the SN contribute negligible $V \& R$-band flux to the pre- and post-explosion images respectively.  However, late-time optical spectra of the SN site presented by \citet{2009ApJ...704..306K}, and discussed in more detail in $\S$\ref{sec:echo}, clearly show broad emission lines that can only be attributed to the SN.  Hence we know that SN 2004et {\em does} contribute at least some of the $V \& R$ flux in the late-time WHT images.  These observations therefore point to the third scenario.

\subsubsection{B-band photometry}
In the $B$-band observations it appears that the Li05 candidate may be brighter at late-time; $B_{pre}$ = 24.02$\pm$0.21, $B_{post}$ = 23.62$\pm$0.06.  If the difference is real we would assume that the SN contributes {\em at least} the excess post-explosion flux ($B = 24.9^{+0.8}_{-0.5}$) with the remainder coming from neighbouring unresolved sources, or {\em at most} the entire post-explosion flux.  We predicted the $B$ magnitude of the SN at the time of the WHT observations ($\sim$ 3 yrs post-explosion) to be $\approx$24.7 from extrapolation of the SN light curve in \citet{2010MNRAS.404..981M}.  This extrapolated magnitude is in good agreement with what we measure for the post-explosion excess, suggesting that 1) the SN is still visible, at least in $B$, and 2) the majority of the late-time flux comes from other unresolved sources that would also have been visible pre-explosion.  These observations are in good agreement with what we infer from the $V \& R$-band photometry above.

\subsubsection{I-band photometry}
\label{sec:I_phot}

Photometry of the I-band observations shows that the pre-explosion source coincident with the Li05 candidate is significantly brighter than its late-time counterpart; $I_{pre} = 21.27\pm0.08$, $I_{post} = 21.99\pm0.06$.  Comparing this object to the surrounding stars, it is easy to see from Figure~\ref{fig:pre-post} that it is visibly brighter in the INT image than in the WHT frame.  Such an excess in the pre-explosion image can logically be interpreted as a detection of the SN progenitor, which must have been intrinsically very red or heavily extinguished (or both) to have avoided detection in the $BV \& R$-bands.  An $I$-band magnitude for the unresolved progenitor can be trivially calculated from the difference between the pre- and post-explosion photometry; $I_{prog} = 22.06 \pm 0.20$.  

We predicted the magnitude of the SN at the epoch of the WHT observations to be $I\approx$24.6, by extrapolating the $I$-band light curve of \citet{2010MNRAS.404..981M}.  If correct, the predicted photometry would suggest that the SN contributes $\sim$9 per cent of the WHT $I$-band flux, which should be discounted when calculating the progenitor magnitude.  This would result in a slightly brighter progenitor magnitude of $I_{prog} = 21.96 \pm 0.20$.  However, in extrapolating the $BVR\,\&\,I$ SN photometry we have assumed that there was no flattening or re-brightening of these light curves due to light echoes or ejecta-CSM (circumstellar material) interaction.  \citet{2009ApJ...704..306K} found that, starting October 2007, SN~2004et {\em did} in fact re-brighten significantly in mid-IR Spitzer data.  However, previous mid-IR observations from August 2007 (around the time of our WHT observations) had still shown a declining light curve \citep{2009ApJ...704..306K}, so our assumption above may not be unreasonable.  We will discuss the IR-echo in more detail in $\S$\ref{sec:echo}.  Later in this paper ($\S$\ref{sec:04et_final_prog_properties}) we attempt to use the late-time optical spectra of \citet{2009ApJ...704..306K} to constrain the true contribution of the SN to the late-time optical photometry, and subsequently to correct the progenitor magnitudes.

\begin{table}
\caption{Ground-based pre-explosion and late-time photometry of the source close to the position of SN 2004et}
\begin{center}
\begin{tabular}{lccc}
\hline\hline
Filter & Pre-SN CFHT & Pre-SN CFHT \& INT & Post-SN WHT \\
      & (Li et al. 2005) & (this work) &        \\
\hline
B & 24.30(0.48) & 24.02(0.21) & 23.62(0.06) \\
V & 22.93(0.31) & 23.04(0.27) & 23.06(0.04)\\
R & 22.50(0.18) & 22.24(0.06) & 22.27(0.05)\\
I & - & 21.27(0.08) & 21.99(0.06)\\
\hline\hline
\end{tabular}
\end{center}
\label{tab:prepost_phot}
\end{table}

\subsection{High-resolution data: alignment and photometry}
\label{sec:high_res_04et}

\begin{figure*}
    \centering
    \includegraphics[width=160mm]{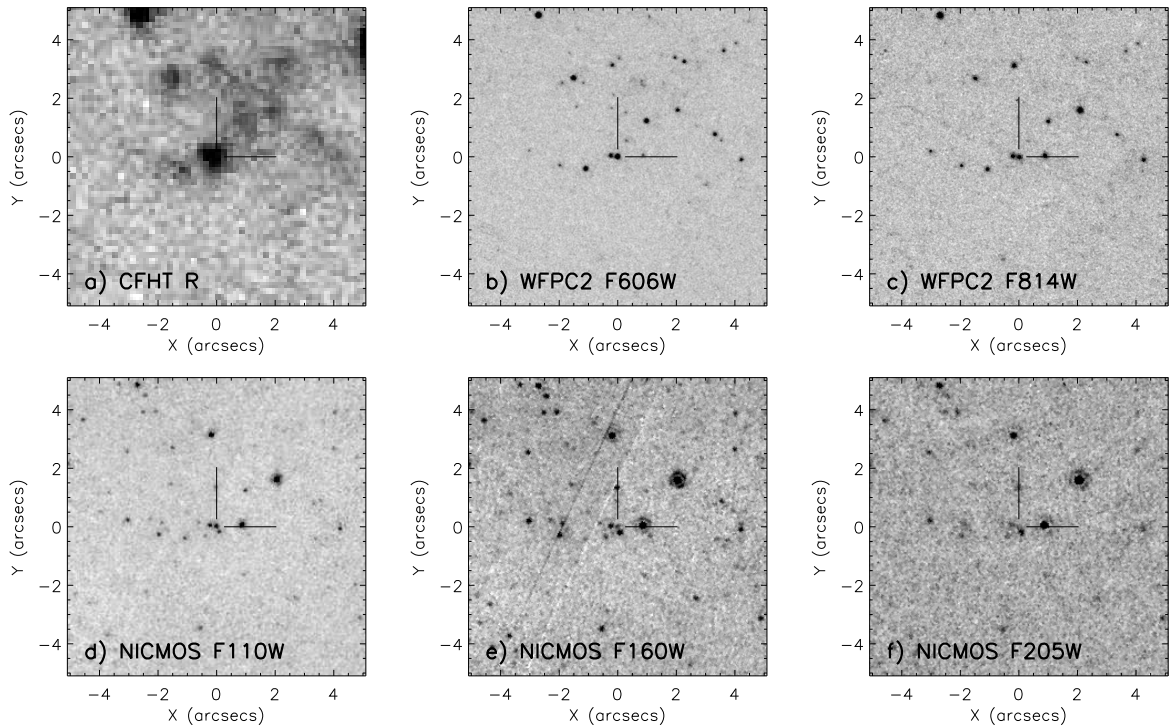}
    \caption{Each panel in this figure is centred on the SN position and oriented such that North is up and East is to the left.  {\bf (a)}: Pre-CFHT $R$-band, {\bf (b\&c)}: WFPC2 F606W and F814W and {\bf (d-f)}: NICMOS F110W, F160W and F205W late time observations taken almost 3 years post-explosion (2007 July 08).  The single source close to the SN position in the ground-based image is resolved into at least 3 objects in the HST images (see Table~\ref{tab:hst_late_phot} for photometry). One of these objects (indicated by the cross hairs) is exactly coincident with the SN position, but it is unclear if it is a detection of the SN alone.}
\label{fig:hst_latetime}
\end{figure*}

Immediately apparent from the HST and Gemini AO images is that the previously unresolved source, coincident with the Li05 candidate progenitor, in fact consists of at least three distinct sources (see Figures~\ref{fig:hst_latetime}\&\ref{fig:NIR_echo}).  This provides conclusive evidence that the Li05 candidate identified in ground-based, seeing limited observations, was {\em not solely} the progenitor of SN 2004et.  What contribution the unresolved progenitor made to the pre-explosion images is the question we attempt to address in the remainder of $\S$\ref{sec:2004et}.

Photometry of the WFPC2 and NICMOS datasets was carried out using PSF fitting techniques within {\sc iraf daophot}.  Empirical PSF models were created using several isolated stars in each image.  The updated WFPC2 zeropoints of \citet{Dolphin:2000p1141} were taken from Andrew Dolphin's website\footnote{http://purcell.as.arizona.edu/wfpc2\_calib}, while NICMOS Vegamag zeropoints were determined using data from the STScI site\footnote{http://www.stsci.edu/hst/nicmos/performance/photometry}.  Flight system magnitudes measured for the three sources clustered close to the SN site are recorded in Table~\ref{tab:hst_late_phot}.  

\begin{table*}
\caption{WFPC2 and NICMOS late-time Vegamag photometry of resolved sources close to SN site.  Each source is labelled with reference to its position relative to the SN position (see Figure~\ref{fig:hst_latetime}).  Note the source labelled ``Centre'' is exactly coincident with the position of SN~2004et.}
\begin{center}
\begin{tabular}{lccccc}
\hline\hline
\multicolumn{6}{r}{{\underline{WFPC2}\,\,\,\,\,\,\,\,\,\,\,\,\,\,\,\,\,\,\,\,\,\,\,\,\,\,\,\,\,\,\,\,\,\,\,\,\,\,\,\,\,\,\,\,\,\,\,\,\,\,\,\,\,\,\,\,\,\,\,\,\,\,\,\,\,\,\,\,\,\,\,\underline{NICMOS}\,\,\,\,\,\,\,\,\,\,\,\,\,\,\,\,\,\,\,\,\,\,\,\,\,\,\,\,\,\,\,\,\,\,\,\,\,\,\,\,}} \\
\underline{Object} & F606W & F814W & F110W & F160W & F205W\\
Date&&&&&\\
\hline
\underline{East}&&&&&\\
2007 Jul 08 & 23.88(0.04)  & 22.98(0.05)  & 22.52(0.06)  & 21.67(0.05)  & 21.26(0.08)\\
2008 Jan 19 & 23.90(0.03)  & 23.20(0.04)  & 22.64(0.06)  & 21.95(0.06)  & $>$22.1\\  
\underline{Centre}&&&&&\\
2007 Jul 08 & 23.07(0.03)  & 22.80(0.04)  & 22.17(0.06)  & 22.65(0.09)  & 21.70(0.15)\\
2008 Jan 19& 23.29(0.03)  & 23.06(0.04)  & 22.42(0.06)  & 22.56(0.10)  & 21.22(0.09)\\
\underline{South}&&&&&\\
2007 Jul 08 & $>$26.0  & 25.33(0.14)  & 22.86(0.07)  & 21.22(0.04)  & 20.78(0.05)\\
2008 Jan 19& $>$26.3  & 25.16(0.14)  & 22.89(0.07)  & 21.20(0.04)  & 20.75(0.05)\\
\hline\hline
\end{tabular}
\end{center}
\label{tab:hst_late_phot}
\end{table*}

Alignment with the 2005 ACS/HRC F625W image of the SN confirmed that one of the three sources is exactly coincident with SN~2004et, having a measured separation of $3 \pm 8$ mas.  This object is labelled ``Centre'' in Table~\ref{tab:hst_late_phot} and is indicated by the cross hairs in Figure~\ref{fig:hst_latetime}.  However, with the current dataset it is impossible to determine whether object Centre is solely a detection of the SN in all filters, or whether a second unresolved source (e.g. a close companion of the former progenitor star) is now also visible.

What does appear clear is that object Centre is significantly brighter than one would expect the SN to be from simple extrapolation of the late-time light curves of \citet{2010MNRAS.404..981M}.  At the time of the 2007 July 08 HST observations, we estimated the SN to have magnitudes of $V\approx$25.7 and $I\approx$24.2.  This compares with object Centre magnitudes of $F606W=23.07\pm0.03$ ($V$-band = 23.18) and $F814W=22.80\pm0.03$ ($I$-band = 22.78).\footnote{HST photometry transformed to standard Johnson-Cousins $V$ \& $I$ magnitudes using the colour transformations from http://purcell.as.arizona.edu/wfpc2\_calib}  

However, we know from \citet{2009ApJ...704..306K} that the SN re-brightened significantly in the mid-IR beginning in October 2007.  Although no mid-IR re-brightening was detected earlier than this date (including observations from August 2007) we cannot completely rule out the possibility of an optical light-echo, which could significantly increase the contribution of the SN to object Centre in the HST image from 2007 July 08.  It is therefore also entirely possible that object Centre {\em is} solely a detection of SN 2004et.  We consider this case in $\S$\ref{sec:all_SN}, while in $\S$\ref{sec:04et_final_prog_properties} we use the late-time optical spectra of the SN site from \citet{2009ApJ...704..306K} to constrain the minimum contribution of the SN to the late-time optical photometry.

\subsubsection{NIR re-brightening of SN 2004et and late-time optical spectra}
\label{sec:echo}

\begin{figure*}
    \centering
    \includegraphics[width=160mm]{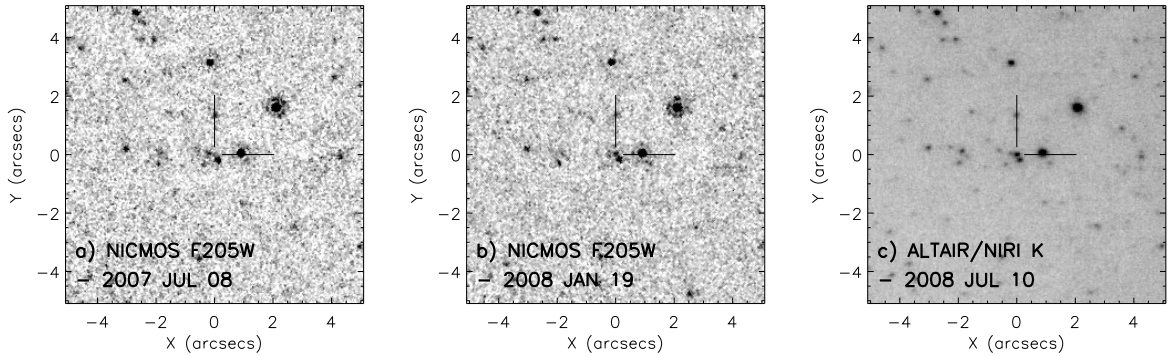}
    \caption[]{Comparison of {\bf (a)}: NICMOS F205W image of the field of SN 2004et from 2007 July 08, {\bf (b)}: NICMOS F205W image from 2008 Jan 08, and {\bf (c)}: Gemini Altair/NIRI $K$ image from 2008 July 10.  The images are centred on the SN position, which is also marked by cross hairs.  The HST and Gemini images from 2008 clearly shows that SN 2004et has re-brightening significantly in F205W / $K$.}
\label{fig:NIR_echo}
\end{figure*}

Although there is no direct evidence from the HST data for an optical light-echo associated with SN 2004et, we do observe a significant brightening of object Centre in the near-IR (NIR), both in the NICMOS F205W and Gemini Altair/NIRI $K$-band observations from 2008 Jan 19 and 2008 Jul 10 respectively (see Figure~\ref{fig:NIR_echo}).  Object Centre, which is marginally detected in the F205W observation from 2007 July 08, increases in brightness by roughly 40 per cent by 2008 Jan 19.  Photometry of the Gemini adaptive-optics observations was complicated by the large variation in the PSF across the FOV (due to anisoplanatic effects of the AO correction) and by having only one usable standard star (2MASS 20352132+6007102).  Nevertheless, we estimated an approximate magnitude for object Centre in the 2008 July 10 Gemini image of $K \approx$ 20.0.

We attribute the increase in object Centre's F205W / $K$-band flux to a re-brightening of SN 2004et in the NIR, but stress once again that we cannot be certain that object Centre consists solely of the SN.  The NIR re-brightening of SN 2004et was contemporaneous with the mid-IR increase observed by \citet{2009ApJ...704..306K}.

Interestingly, we do not observe a brightening of object Centre in any of the other optical or NIR observations shortward of $\sim2\mu$m.  However, we note that the rate of decline in the optical bands between July 2007 and January 2008 is much slower than that measured by \citet{2010MNRAS.404..981M} for the earlier part of the radioactive tail; $\sim0.15$ mag/100 days for the late-time HST photometry compared to $\sim1.1$ mag/100 days from Maguire et al.  Furthermore, the late-time F160W photometry show no discernible decline whatsoever.  All this could be interpreted as a {\em flattening} of the SN 2004et light curves caused by the same mechanism(s) that produced the F205W / $K$ and mid-IR re-brightening.  Another possible explanation is that a second unresolved source at the position of object Centre (such as a close companion star to the former SN progenitor) is beginning to dominate over the fading SN.  A combination of the two effects is also possible.  Future HST observations of the SN site will be required to determine which is actually the case.

Of immediate consequence for our study of the progenitor star in the {\em optical}, pre-explosion images are the late-time, {\em optical} spectroscopic observations of the site of SN 2004et presented in \citet{2009ApJ...704..306K}, and reproduced here (Figure~\ref{fig:04et_keckspec}) with the permission of the author.  The three spectra, taken with the Keck Telescope, span the period between $\sim$2 to 3 yrs post-explosion, and clearly show broad emission lines that can only be attributed to the SN.  The width of these lines indicate expansion velocities of $\sim8500\,km\,s^{-1}$, and Kotak et al. interpret this as being due to the impact of the SN ejecta on the progenitor circumstellar medium.  Such interaction would have resulted in a slower rate of decline ({\em flattening}) of the optical light curves.  This lends support to the first scenario in the previous paragraph, but lacking further observations we still cannot rule out the possibility of a second unresolved source at the SN position.  

The Keck spectra are certainly contaminated by flux from other resolved sources close to the SN site.  The spectra were reportedly taken under relatively poor conditions using a slit width of around 1.5$\arcsec$ (R. Kotak, private communication).  A glance at Figure~\ref{fig:hst_latetime} of this paper shows that, depending on the slit orientation, several objects could have contributed significant flux, not least the object immediately to the east of the SN position.  It is therefore impossible to determine the {\em total} flux contribution made by SN 2004et to the late-time optical spectra / photometry.

\subsection{Progenitor luminosity and mass estimates}
\label{sec:04et_mass}

In this section we attempt to derive luminosity and mass estimates for the progenitor of SN 2004et.  In $\S$\ref{sec:I_phot} we detected an unresolved progenitor star as an $I$-band flux excess in pre-explosion INT observations, when these were compared with WHT images taken $\sim$3 yrs post-explosion.  The optical spectra of \citet{2009ApJ...704..306K} (Fig.~\ref{fig:04et_keckspec}) show that the SN {\em was} still visible at this late-epoch, and hence we must correct for its contribution to the late-time optical photometry before calculating the magnitude of the unresolved progenitor star. 

As discussed above, determination of the {\em total} flux contribution of the SN to the late-time photometry is impossible.  The Keck spectra are contaminated by nearby, resolved sources, while object Centre in the HST observations may consist of more than just SN 2004et.  We can, however, estimate the {\em maximum} and {\em minimum} contributions of the SN from the HST photometry and the Keck spectra.  These two extremes are considered separately in the following subsections, where we also determine corresponding values for the progenitor photometry, luminosity and mass.

We also note that there is some evidence from the photometry in Table~\ref{tab:hst_late_phot} to suggest that object East may be variable.  This adds an unknown systematic uncertainty to our calculations in the following subsections.  Further HST observations of the SN site will be required to determine if object East is truly variable.

To calculate the progenitor luminosity, and ultimately estimate its mass from stellar models, we require the distance to the host galaxy, the extinction along the line of sight and the (approximate) metallicity of the progenitor star.  \citet{2009MNRAS.398.1041B} present a review of the distance estimates to NGC 6946 in the literature, and employ the unweighted mean as their preferred value; $\mu$ = 28.78$\pm$0.08 (d = 5.7$\pm$0.2 Mpc).  We adopt the same value in this paper.  

The extinction of \mbox{E$(B\!-\!V)$} = $0.41\pm0.07$ was estimated by \citet{Zwitter:2004p1930} from the equivalent width of the Na {\sc i} D lines in a high resolution spectrum of SN 2004et, and using the calibration of \citet{Munari:1997p1098}.  The reddening uncertainty was chosen by Li05 to bracket the lower limit, which assumes no host-galaxy extinction.  As pointed out by Li05, the Galactic component of extinction towards SN~2004et is estimated to be \mbox{E$(B\!-\!V)$} = 0.34 \citep{Schlegel:1998p1933}.  

We also attempted to constrain the metallicity of the progenitor of SN~2004et from the available literature.  \citet{Pilyugin:2004p1923} studied 9 H\,{\sc ii} regions in NGC 6946, but none of these are close to the location of SN.  We therefore used their calculated abundance gradient and the de-projected galactocentric radius of SN~2004et ($R_{G}/R_{25}$ = 0.92) to estimate the metallicity at its position.  This yielded an oxygen abundance of 12+log(O/H) = 8.3 dex, which is consistent with that of the LMC \citep{Hunter:2007p1883}.

\subsubsection{Object Centre is SN 2004et}
\label{sec:all_SN}

Firstly, we consider the {\em maximum} possible flux contribution of the SN to the late-time photometry.  This is the case where SN 2004et is assumed to contribute {\em all} the flux of object Centre in the late-time HST images (Figure~\ref{fig:hst_latetime} \& Table~\ref{tab:hst_late_phot}).  In this scenario, the pre-explosion source (Li05 candidate) is logically assumed to consist of the other HST resolved objects  (i.e. objects East and South) plus the SN progenitor.  By subtracting the flux of objects East and South from the ground-based, pre-explosion photometry, we can calculate photometry for the progenitor star.  To this end, the July 2007 $F606W$ and $F814W$ photometry of object East was transformed to standard Johnson-Cousins $V$ \& $I$ magnitudes using the colour transformations from Andrew Dolphin's website$^{11}$.  With no detection of object South in $F606W$, we were unable to transform its $F814W$ photometry to the standard system.  However, we note that, in the case of the $F814W$-to-$I$ transformation, the correction is generally very small.  Johnson-Cousins photometry for object East was therefore calculated as $V$ = 24.25$\pm$0.03 and $I$ = 22.94$\pm$0.04, while for object South we set $I$ = $F814W$ = 25.33$\pm$0.14.

These magnitudes and their respective errors were converted to flux units, and subtracted from the $V$ \& $I$-band photometry of the pre-explosion source (Table~\ref{tab:prepost_phot}).  This yielded progenitor star magnitudes of, $V_{prog}$ = 23.47$\pm$0.40 and $I_{prog}$ = 21.57$\pm$0.12.  The apparent colour of the progenitor was therefore {\mbox{$(V\!-\!I)$}} $ = 1.90\pm0.41$.  Applying our best estimate of extinction (see $\S$\ref{sec:04et_mass}), we calculated the progenitor's intrinsic colour; {\mbox{$(V\!-\!I)_o$}} $ = 1.24\pm0.48$.  This colour range corresponds to supergiant stars of type F8 to K3, which are significantly hotter than the M-type red supergiant one would expect for a type II-P SN progenitor.  \citet{2000asqu.book..381D} show that the values of {\mbox{$(V\!-\!I)$}} + BC for such stars are quite similar, and we calculated an average of {\mbox{$(V\!-\!I)$}} + BC = 0.75$\pm$0.15.  Applying this to the $I$-band detection, along with values for the extinction and distance modulus detailed above ($\S$\ref{sec:04et_mass}), we calculated the progenitor's bolometric magnitude, M$_{\rm bol} = -7.06\pm0.23$ and its luminosity, \logl = 4.72$\pm$0.09.  By comparing this luminosity to the STARS stellar models of LMC metallicity in Figure~\ref{fig:STARS}, we estimated the progenitor initial mass to be $10^{+5}_{-1}$ M$_{\odot}$.

\subsubsection{Estimating the minimum contribution of SN 2004et from late-time optical spectra}
\label{sec:04et_final_prog_properties}

In this section we have attempted to estimate the {\em minimum} flux contribution of the SN to the late-time photometry.  In order to constrain the SN flux, we firstly utilised the {\sc stsdas synphot} package to measure the SN emission line fluxes in the late-time optical spectra from \citet{2009ApJ...704..306K}.  By interpolating the measurements from the 2007 April 13 and 2007 November 12 spectra we derived emission line magnitudes that were contemporaneous with our 2007 August 12 WHT photometry, assuming an exponential decay in flux.  The reader should note that these ground-based spectroscopic observations were made under natural seeing conditions, and hence the continuum flux contains contributions from at least three different sources, that is those resolved in the HST and Gemini AO observations.  

\begin{figure}
   \centering
   \includegraphics[width=80mm]{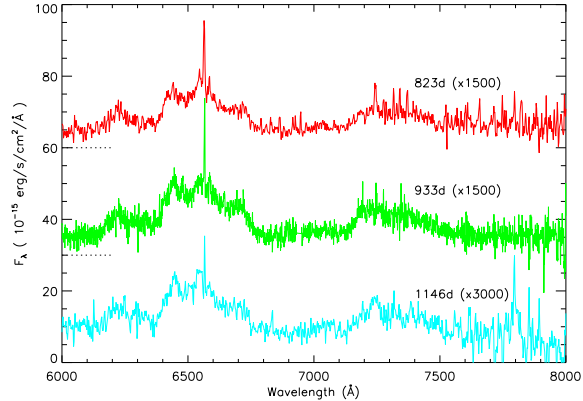}
   \caption[]{Late-time optical spectra of SN 2004et from \citet{2009ApJ...704..306K}, taken at the Keck telescope. Fluxes are as presented by Kotak et al.  The spectra have been scaled by the amounts shown in brackets.  The upper two spectra have been shifted vertically for clarity.  The fluxes are inconsistent with our WHT and HST late-time photometry and we estimate that the fluxes shown here are too large by approximately a factor of 2, at least for the 933d and 1146d spectra.  The broad emission lines, particularly H$_{\alpha}$, are clearly visible.  Courtesy of \citet{2009ApJ...704..306K}}
\label{fig:04et_keckspec}
\end{figure}

Magnitudes were first of all measured from the flux-calibrated spectra as presented in \citet{2009ApJ...704..306K} (Figure~\ref{fig:04et_keckspec}), but the interpolated results, $V_{spec}$ = 22.22 and $R_{spec}$= 21.59, were found to be significantly brighter than our WHT photometry (see Table~\ref{tab:prepost_phot}).  $V_{spec}$ and $R_{spec}$ were brighter by factors of $\times$2.2 and $\times$1.9 respectively.  Since we do not know the exact conditions under which the spectra were taken, such as seeing and slit orientation, we cannot easily quantify the contamination due to other nearby resolved sources (see Figure~\ref{fig:pre-post}).  In the $R$-band, at least, the blended source closest to the SN site is much brighter than any of the nearby objects.  We suggest that the neighbouring resolved sources could not have resulted in the observed $R$-band excess in the spectra, and subsequently conclude that the flux values reported by \citet{2009ApJ...704..306K} are approximately a factor of 2 higher than their true values.  In the analysis that follows we have re-scaled the spectra by a factor of $\times$0.5, so as to be consistent with our WHT photometry.

In order to estimate the the emission line magnitudes, the continuum of each spectrum was fitted and removed before the {\sc synphot} task {\em calcphot} was run using the Landolt $V$ \& $R$ passbands.  Since {\em calcphot} resets all negative fluxes to zero, positive noise spikes lead to artificial brightening of the measured magnitudes.  We therefore added a constant flux offset to the continuum subtracted spectra before running {\em calcphot}, which ensured that all fluxes were greater than zero.  The offset magnitudes were measured separately, and then subtracted from the combined results to find the true emission line magnitudes.  Interpolation of these results yielded estimates of the emission line magnitudes on the date of the WHT observation; 2007 August 12.  Following this procedure we found that the SN emission lines could account for just 5 per cent of the total flux in $V$, but around 25 per cent of that in the $R$-band.  This is not surprising, since the $R$ filter covers the wavelength region that includes the prominent H$_{\alpha}$ and [Ca II] emission features, while the $V$-band only grazes the H$_{\alpha}$ profile.

In the $I$-band wavelength range the spectra from \citet{2009ApJ...704..306K} were noisy and included prominent sky lines.  However, the  [Ca II] emission feature lies within the bounds of this passband.  We estimated its contribution to the WHT $I$-band observations in the same manner as described for the other filters above, except that in this case all flux values at wavelengths longer than 7600\AA\ were set to zero in order to suppress the sky lines.  The [Ca II] emission was thereby estimated to contribute around 9 per cent of the total flux in $I$; a lower limit to the SN emission line flux in this filter.

Of course, we expect the SN to have contributed flux to the continuum also, although the absolute level of this contribution is impossible to measure from the current dataset.  By comparing the $V$ \& $R$ magnitudes of the original 2007 April 13 and 2007 November 12 spectra, we found that around 14-18 per cent of the April flux is lost from the continuum by the time of the November observation.  These continuum losses equate to roughly 9$\times$10$^{-4}$ mag\,d$^{-1}$.  Between our WHT observations on 2007 August 12, and the final spectrum on 2007 November 12, we calculate that the SN continuum would have faded by $\sim$0.08 mag in $V$ \& $R$, again assuming an exponential decay in flux.  This constitutes a lower limit for the continuum contribution of the SN in the WHT observations.  Later spectra may well have revealed a further decline in the continuum due to the fading SN.  However, lacking such extra information, we only consider flux that can be logically attributed to the SN using the current data, and reiterate that it is a {\em minimum} value.  (The reader should note that the combined emission-line plus continuum contributions derived here are in excess of the decline observed between the 2007 July and 2008 January HST optical observations.  Hence, no further constraint can be gleaned from the HST data).

Our estimates of the {\em minimum} SN flux in $V,R \& I$ were subtracted from the WHT post-explosion photometry of the Li05 candidate in order to calculate the contribution from the other unresolved stars.  Subtracting the result from the pre-explosion photometry of Li05 candidate left us with the flux of the progenitor star itself.  In this way we estimated the magnitudes of the progenitor in the pre-explosion images to be $V_{prog}$=25.13$^{+\infty}_{-1.2}$, $R_{prog}$=23.39$\pm$0.26 and $I_{prog}$=21.88$\pm$0.17. (We assume a similar continuum reduction in $I$ as we have observed in $V$ \& $R$).  Clearly, the calculated flux of $V_{prog}$ is insignificant when compared to its error.  Therefore, we use only $R_{prog}$ and $I_{prog}$ to characterise the progenitor in this case.  Using the extinction quoted in $\S$\ref{sec:04et_mass}, the intrinsic colour of the progenitor is calculated as {\mbox{$(R\!-\!I)_o$}} $ = 1.16\pm0.36$.  These colours correspond to supergiant stars of types K4 to M4.  Again, we noted that the {\mbox{$(V\!-\!I)$}} + BC values for these stars are quite similar, and calculated an average value of 0.89$\pm$0.11 from \citet{Elias:1985p1934} and \citet{2000asqu.book..381D}.  The bolometric magnitude and luminosity were calculated from $I_{prog}$ to be M$_{\rm bol} = -6.62\pm0.24$, and \logl = 4.54$\pm$0.10.  This corresponds to an initial mass for the progenitor of $8^{+5}_{-1}$ M$_{\odot}$.  

Having corrected for the emission line fluxes and minimum continuum contributions of the SN, we conclude that these results are our best estimates for the spectral type, luminosity and initial mass of the progenitor of SN 2004et.  However, we stress that future high-resolution observations of the SN site using HST will be required to determine if object Centre has faded further in order to confirm or adjust these results.

\section{Conclusions}

\begin{table*}
\caption{Summary of photometry, luminosity and mass estimates for the progenitors of SNe 2006my, 2006ov and 2004et}
\begin{center}
\begin{tabular}{lccccccc}
\hline\hline
Supernova & Photometry & Distance Modulus & Extinction & Colour & Spectral Type & Luminosity & Mass\\
   & & & {\mbox{$E(B-V)$}} & & & \logl & M$_{\odot}$       \\
\hline
2006my & $F814W > 24.8$ & $31.74\pm0.25$ & 0.027 & - & M0\,I (assumed) & $< 4.51$ & $< 13$\\
&&&&\\
2006ov & $F814W > 24.2$ & $30.50\pm0.40$ & 0.022 & - & M0\,I (assumed) & $< 4.29$ & $< 10$\\
&&&&\\
2004et (a) & $I = 21.57\pm0.12$ & $28.78\pm0.08$ & $0.41\pm0.07$ & {\mbox{$(V\!-\!I)_o$}} = $1.24\pm0.48$ & F8\,I to K3\,I & $4.72\pm0.09$ & $10^{+5}_{-1}$\\
\ \ \ \ \ \ \ \ \ \ \ \,(b) & $I = 21.88\pm0.17$ & $28.78\pm0.08$ & $0.41\pm0.07$ & {\mbox{$(R\!-\!I)_o$}} = $1.16\pm0.36$ & K4\,I to M4\,I & $4.54\pm0.10$ &  $8^{+5}_{-1}$\\
\hline\hline
\multicolumn{7}{l}{(a) assuming SN 2004et contributes $all$ the flux of object Centre in late time HST images ($\S$\ref{sec:all_SN})}\\
\multicolumn{7}{l}{(b) SN 2004et contribution to object Centre estimated from late-time optical spectra ($\S$\ref{sec:04et_final_prog_properties})}
\end{tabular}
\end{center}
\label{tab:results_summary}
\end{table*}

We have re-analysed the pre-explosion observations of the type II-P SNe 2006my, 2006ov and 2004et.  In the cases of SNe 2006my and 2006ov we have argued that the published progenitor candidates (Li07) are most likely {\em not} coincident with their respective supernova sites in pre-explosion HST+WFPC2 observations.  We concluded that the progenitors of SNe 2006my and 2006ov were probably not detected, and hence derived detection limits for each (see Table~\ref{tab:results_summary}).  Based on the assumption that the unseen progenitors were red supergiants, a reasonable assumption given both SNe were classified as type II-P \citep{1976ApJ...207..872C,Arnett:1980p1887}, we calculated upper luminosity and mass limits for the progenitors of SNe 2006my ($\log L/L_{\odot} < 4.51$; $m < 13$M$_{\odot}$) and 2006ov ($\log L/L_{\odot} < 4.29$; $m < 10$M$_{\odot}$).  Future high-resolution observations of each SN site will be required to confirm these results.

Late-time WHT observations of the site of SN~2004et revealed that the {\em yellow} supergiant progenitor candidate, identified in pre-explosion CFHT $BVR$ data by Li05, was still visible some 3 years post-explosion.  High-resolution, late-time Gemini AO and HST images resolve this object into at least 3 distinct sources, one of which (object Centre) is coincident with the SN.  A hitherto unpublished, pre-explosion $i'$-band image from the INT was also analysed.  Through comparison with the late-time WHT $I$-band data, we discovered a pre-explosion excess of flux coincident with the Li05 progenitor candidate.  We attributed this to a detection of the unresolved progenitor star.

Gemini Altair/NIRI and HST NICMOS observations revealed an apparent re-brightening of the SN in the NIR $\sim$3.5 - 4 years post-explosion.  This coincides with the re-brightening observed by Kotak et al. (2009) in the mid-IR.  Late-time optical spectra presented by Kotak et al. confirmed that SN 2004et was still visible in the optical at the time of our WHT photometry.  Correcting for the SN contribution we found an excess of pre-explosion flux in the $R$-band, which we again attributed to the unresolved progenitor.  By combining the $R$ and $I$-band progenitor photometry we estimated the progenitor of SN 2004et to be a K4 to M4 supergiant, of \logl = 4.54$\pm$0.10, and initial mass $8^{+5}_{-1}$ M$_{\odot}$ (see Table~\ref{tab:results_summary}).  These constitute our best estimates for the spectral type, luminosity and mass of the progenitor of SN 2004et, but we stress that future observations may show that the SN has faded still further, in which case our progenitor photometry would require revision.

\section*{Acknowledgments}

We thank an anonymous referee for detailed comments and suggestions which substantially improved the paper.

This research is based in part on observations made with the NASA/ESA Hubble Space Telescope, obtained from the data archive at the Space Telescope Science Institute. Support for Program number 10803 was provided by NASA through a grant from STScI, which is operated by the Association of Universities for Research in Astronomy, Incorporated, under NASA contract NAS 5-26555.

The William Herschel Telescope and Isaac Newton Telescope are operated on the island of La Palma by the Isaac Newton Group in the Spanish Observatorio del Roque de los Muchachos of the Instituto de Astrof'sica de Canarias.  We thank the Service Programme and Ian Hunter for coordination of the WHT observations.  The INT data, on which this research is partly based, was made publicly available through the Isaac Newton Group's Wide Field Camera Survey Programme. 
      
This research used the facilities of the Canadian Astronomy Data Centre operated by the National Research Council of Canada with the support of the Canadian Space Agency, and is partly based on observations obtained at the Canada-France-Hawaii Telescope (CFHT) which is operated by the National Research Council of Canada, the Institut National des Sciences de l'Univers of the Centre National de la Recherche Scientifique of France,  and the University of Hawaii.

Based on observations obtained at the Gemini Observatory (acquired through the Gemini Science Archive), which is operated by the Association of Universities for Research in Astronomy, Inc., under a cooperative agreement with the NSF on behalf of the Gemini partnership: the National Science Foundation (United States), the Particle Physics and Astronomy Research Council (United Kingdom), the National Research Council (Canada), CONICYT (Chile), the Australian Research Council (Australia), CNPq (Brazil) and SECYT (Argentina). 

This work conducted as part of the award ``Understanding the lives of massive stars from birth to supernovae'' (S. J. Smartt) made under the European Heads of Research Councils and European Science Foundation EURYI (European Young Investigator) Awards Scheme, was supported by funds from the Participating Organisations of EURYI and the EC Sixth Framework Programme.  SJS and RMC thank the Leverhulme Trust and the European Science Foundation for a Philip Leverhulme Prize and postgraduate funding.  S.M. acknowledges support from the Academy of Finland (project 8120503).

\bibliography{bibtex.bib}

\begin{thebibliography}{72}
\expandafter\ifx\csname natexlab\endcsname\relax\def\natexlab#1{#1}\fi

\bibitem[{{Anderson} \& {King}(2003)}]{2003PASP..115..113A}
{Anderson}, J. \& {King}, I.~R. 2003, \pasp, 115, 113

\bibitem[{Argo {et~al.}(2005)Argo, Beswick, Muxlow, Pedlar, Fenech, \&
  Thrall}]{Argo:2005p1792}
Argo, M.~K., Beswick, R.~J., Muxlow, T. W.~B., {et~al.} 2005, Memorie della
  Societa Astronomica Italiana, 76, 565

\bibitem[{Arnett(1980)}]{Arnett:1980p1887}
Arnett, W.~D. 1980, ApJ, 237, 541

\bibitem[{Asplund {et~al.}(2004)Asplund, Grevesse, Sauval, Prieto, \&
  Kiselman}]{Asplund:2004p1946}
Asplund, M., Grevesse, N., Sauval, A.~J., Prieto, C.~A., \& Kiselman, D. 2004,
  A\&A, 417, 751

\bibitem[{{Bethe}(1990)}]{1990RvMP...62..801B}
{Bethe}, H.~A. 1990, Reviews of Modern Physics, 62, 801

\bibitem[{Blondin {et~al.}(2006)Blondin, Modjaz, Kirshner, \&
  Challis}]{Blondin:2006p705}
Blondin, S., Modjaz, M., Kirshner, R., \& Challis, P. 2006, CBET

\bibitem[{{Botticella} {et~al.}(2009){Botticella}, {Pastorello}, {Smartt},
  {Meikle}, {Benetti}, {Kotak}, {Cappellaro}, {Crockett}, {Mattila}, {Sereno},
  {Patat}, {Tsvetkov}, {van Loon}, {Abraham}, {Agnoletto}, {Arbour}, {Benn},
  {di Rico}, {Elias-Rosa}, {Gorshanov}, {Harutyunyan}, {Hunter}, {Lorenzi},
  {Keenan}, {Maguire}, {Mendez}, {Mobberley}, {Navasardyan}, {Ries},
  {Stanishev}, {Taubenberger}, {Trundle}, {Turatto}, \&
  {Volkov}}]{2009MNRAS.398.1041B}
{Botticella}, M.~T., {Pastorello}, A., {Smartt}, S.~J., {et~al.} 2009, \mnras,
  398, 1041

\bibitem[{Cardelli {et~al.}(1989)Cardelli, Clayton, \&
  Mathis}]{Cardelli:1989p1230}
Cardelli, J.~A., Clayton, G.~C., \& Mathis, J.~S. 1989, ApJ, 345, 245

\bibitem[{{Casertano} \& {Wiggs}(2001)}]{Casertano_ISR}
{Casertano}, S. \& {Wiggs}, M. 2001, WFPC2 ISR 01-10, Baltimore, STScI

\bibitem[{{Chevalier}(1976)}]{1976ApJ...207..872C}
{Chevalier}, R.~A. 1976, ApJ, 207, 872

\bibitem[{Chevalier {et~al.}(2006)Chevalier, Fransson, \&
  Nymark}]{Chevalier:2006p1775}
Chevalier, R.~A., Fransson, C., \& Nymark, T.~K. 2006, ApJ, 641, 1029

\bibitem[{Dolphin(2000{\natexlab{a}})}]{Dolphin:2000p1141}
Dolphin, A.~E. 2000{\natexlab{a}}, PASP, 112, 1397

\bibitem[{Dolphin(2000{\natexlab{b}})}]{Dolphin:2000p1123}
Dolphin, A.~E. 2000{\natexlab{b}}, PASP, 112, 1383

\bibitem[{{Drilling} \& {Landolt}(2000)}]{2000asqu.book..381D}
{Drilling}, J.~S. \& {Landolt}, A.~U. 2000, {Normal Stars} (Allen's
  Astrophysical Quantities), 381

\bibitem[{Dwek(1983)}]{Dwek:1983p1936}
Dwek, E. 1983, ApJ, 274, 175

\bibitem[{Eldridge {et~al.}(2007)Eldridge, Mattila, \&
  Smartt}]{Eldridge:2007p556}
Eldridge, J.~J., Mattila, S., \& Smartt, S.~J. 2007, MNRAS: Letters, 376, L52

\bibitem[{Eldridge \& Tout(2004)}]{Eldridge:2004p60}
Eldridge, J.~J. \& Tout, C.~A. 2004, MNRAS, 353, 87

\bibitem[{Elias {et~al.}(1985)Elias, Frogel, \& Humphreys}]{Elias:1985p1934}
Elias, J.~H., Frogel, J.~A., \& Humphreys, R.~M. 1985, ApJ Supplement Series,
  57, 91

\bibitem[{{Elias-Rosa} {et~al.}(2009){Elias-Rosa}, {Van Dyk}, {Li}, {Morrell},
  {Gonzalez}, {Hamuy}, {Filippenko}, {Cuillandre}, {Foley}, \&
  {Smith}}]{2009ApJ...706.1174E}
{Elias-Rosa}, N., {Van Dyk}, S.~D., {Li}, W., {et~al.} 2009, \apj, 706, 1174

\bibitem[{Elmhamdi {et~al.}(2003)Elmhamdi, Danziger, Chugai, Pastorello,
  Turatto, Cappellaro, Altavilla, Benetti, Patat, \&
  Salvo}]{Elmhamdi:2003p1943}
Elmhamdi, A., Danziger, I.~J., Chugai, N., {et~al.} 2003, MNRAS, 338, 939

\bibitem[{Filippenko(1997)}]{Filippenko:1997p259}
Filippenko, A.~V. 1997, Ann. Rev. A\&A, 35, 309

\bibitem[{{Fraser} {et~al.}(2010){Fraser}, {Tak{\'a}ts}, {Pastorello},
  {Smartt}, {Mattila}, {Botticella}, {Valenti}, {Ergon}, {Sollerman}, {Arcavi},
  {Benetti}, {Bufano}, {Crockett}, {Danziger}, {Gal-Yam}, {Maund},
  {Taubenberger}, \& {Turatto}}]{2010ApJ...714L.280F}
{Fraser}, M., {Tak{\'a}ts}, K., {Pastorello}, A., {et~al.} 2010, \apjl, 714,
  L280

\bibitem[{{Fruchter} {et~al.}(2009){Fruchter}, {Sosey}, {Hack}, {Dressel},
  {Koekemoer}, {Mack}, {Mutchler}, \& {Pirzkal}}]{Multidrizzle_handbook}
{Fruchter}, A., {Sosey}, M., {Hack}, W., {et~al.} 2009, "The Multidrizzle
  Handbook", version 3, Baltimore, STScI

\bibitem[{Fruchter \& Hook(2002)}]{Fruchter:2002p1494}
Fruchter, A.~S. \& Hook, R.~N. 2002, PASP, 114, 144

\bibitem[{{Gilmozzi} {et~al.}(1995){Gilmozzi}, {Ewald}, \&
  {Kinney}}]{Gilmozzi_ISR}
{Gilmozzi}, R., {Ewald}, S., \& {Kinney}, E. 1995, WFPC2 ISR 95-02, Baltimore,
  STScI

\bibitem[{Grevesse \& Sauval(1998)}]{Grevesse:1998p1924}
Grevesse, N. \& Sauval, A.~J. 1998, Space Sci. Rev., 85, 161

\bibitem[{Hamuy {et~al.}(2001)Hamuy, Pinto, Maza, Suntzeff, Phillips, Eastman,
  Smith, Corbally, Burstein, Li, Ivanov, Moro-Martin, Strolger, de~Souza, dos
  Anjos, Green, Pickering, Gonz{\'a}lez, Antezana, Wischnjewsky, Galaz, Roth,
  Persson, \& Schommer}]{Hamuy:2001p1942}
Hamuy, M., Pinto, P.~A., Maza, J., {et~al.} 2001, ApJ, 558, 615

\bibitem[{Heger {et~al.}(2003)Heger, Fryer, Woosley, Langer, \&
  Hartmann}]{Heger:2003p40}
Heger, A., Fryer, C.~L., Woosley, S.~E., Langer, N., \& Hartmann, D.~H. 2003,
  ApJ, 591, 288

\bibitem[{Hendry {et~al.}(2005)Hendry, Smartt, Maund, Pastorello, Zampieri,
  Benetti, Turatto, Cappellaro, Meikle, Kotak, Irwin, Jonker, Vermaas,
  Peletier, van Woerden, Exter, Pollacco, Leon, Verley, Benn, \&
  Pignata}]{Hendry:2005p494}
Hendry, M.~A., Smartt, S.~J., Maund, J.~R., {et~al.} 2005, MNRAS, 359, 906

\bibitem[{Holtzman {et~al.}(1995)Holtzman, Hester, Casertano, Trauger, Watson,
  Ballester, Burrows, Clarke, Crisp, Evans, Gallagher, Griffiths, Hoessel,
  Matthews, Mould, Scowen, Stapelfeldt, \& Westphal}]{Holtzman:1995p1198}
Holtzman, J.~A., Hester, J.~J., Casertano, S., {et~al.} 1995, PASP, 107, 156

\bibitem[{Hunter {et~al.}(2007)Hunter, Dufton, Smartt, Ryans, Evans, Lennon,
  Trundle, Hubeny, \& Lanz}]{Hunter:2007p1883}
Hunter, I., Dufton, P.~L., Smartt, S.~J., {et~al.} 2007, A\&A, 466, 277

\bibitem[{{Kotak} {et~al.}(2009){Kotak}, {Meikle}, {Farrah}, {Gerardy},
  {Foley}, {Van Dyk}, {Fransson}, {Lundqvist}, {Sollerman}, {Fesen},
  {Filippenko}, {Mattila}, {Silverman}, {Andersen}, {H{\"o}flich}, {Pozzo}, \&
  {Wheeler}}]{2009ApJ...704..306K}
{Kotak}, R., {Meikle}, W.~P.~S., {Farrah}, D., {et~al.} 2009, \apj, 704, 306

\bibitem[{{Kozhurina-Platais} {et~al.}(2003){Kozhurina-Platais}, {Anderson}, \&
  {Koekemoer}}]{Kozhurina_ISR}
{Kozhurina-Platais}, V., {Anderson}, J., \& {Koekemoer}, A. 2003, WFPC2 ISR
  03-02, Baltimore, STScI

\bibitem[{Leonard {et~al.}(2002)Leonard, Filippenko, Gates, Li, Eastman, Barth,
  Bus, Chornock, Coil, Frink, Grady, Harris, Malkan, Matheson, Quirrenbach, \&
  Treffers}]{Leonard:2002p1941}
Leonard, D.~C., Filippenko, A.~V., Gates, E.~L., {et~al.} 2002, PASP, 114, 35

\bibitem[{Leonard {et~al.}(2008)Leonard, Gal-Yam, Fox, Cameron, Johansson,
  Kraus, Mignant, \& van Dam}]{Leonard:2008p1944}
Leonard, D.~C., Gal-Yam, A., Fox, D.~B., {et~al.} 2008, PASP, 120, 1259

\bibitem[{Levesque {et~al.}(2005)Levesque, Massey, Olsen, Plez, Josselin,
  Maeder, \& Meynet}]{Levesque:2005p1919}
Levesque, E.~M., Massey, P., Olsen, K. A.~G., {et~al.} 2005, ApJ, 628, 973

\bibitem[{Levesque {et~al.}(2006)Levesque, Massey, Olsen, Plez, Meynet, \&
  Maeder}]{Levesque:2006p1920}
Levesque, E.~M., Massey, P., Olsen, K. A.~G., {et~al.} 2006, ApJ, 645, 1102

\bibitem[{Li {et~al.}(2005{\natexlab{a}})Li, Dyk, Filippenko, \&
  Cuillandre}]{Li:2005p1115}
Li, W., Dyk, S. D.~V., Filippenko, A.~V., \& Cuillandre, J.-C.
  2005{\natexlab{a}}, PASP, 117, 121

\bibitem[{Li {et~al.}(2006)Li, Dyk, Filippenko, Cuillandre, Jha, Bloom, Riess,
  \& Livio}]{Li:2006p1886}
Li, W., Dyk, S. D.~V., Filippenko, A.~V., {et~al.} 2006, ApJ, 641, 1060

\bibitem[{Li {et~al.}(2005{\natexlab{b}})Li, Filippenko, \& van
  Dyk}]{Li:2005p1927}
Li, W., Filippenko, A.~V., \& van Dyk, S.~D. 2005{\natexlab{b}}, ATEL, 492, 1

\bibitem[{Li {et~al.}(2007)Li, Wang, Dyk, Cuillandre, Foley, \&
  Filippenko}]{Li:2007p656}
Li, W., Wang, X., Dyk, S. D.~V., {et~al.} 2007, ApJ, 661, 1013

\bibitem[{{Maguire} {et~al.}(2010){Maguire}, {di Carlo}, {Smartt},
  {Pastorello}, {Tsvetkov}, {Benetti}, {Spiro}, {Arkharov}, {Beccari},
  {Botticella}, {Cappellaro}, {Cristallo}, {Dolci}, {Elias-Rosa}, {Fiaschi},
  {Gorshanov}, {Harutyunyan}, {Larionov}, {Navasardyan}, {Pietrinferni},
  {Raimondo}, {di Rico}, {Valenti}, {Valentini}, \&
  {Zampieri}}]{2010MNRAS.404..981M}
{Maguire}, K., {di Carlo}, E., {Smartt}, S.~J., {et~al.} 2010, \mnras, 404, 981

\bibitem[{Mart{\'\i}-Vidal {et~al.}(2007)Mart{\'\i}-Vidal, Marcaide, Alberdi,
  Guirado, Lara, P{\'e}rez-Torres, Ros, Argo, Beswick, Muxlow, Pedlar, Shapiro,
  Stockdale, Sramek, Weiler, \& Vinko}]{MartiVidal:2007p1547}
Mart{\'\i}-Vidal, I., Marcaide, J.~M., Alberdi, A., {et~al.} 2007, A\&A, 470,
  1071

\bibitem[{{Mattila} {et~al.}(2008){Mattila}, {Smartt}, {Eldridge}, {Maund},
  {Crockett}, \& {Danziger}}]{2008ApJ...688L..91M}
{Mattila}, S., {Smartt}, S.~J., {Eldridge}, J.~J., {et~al.} 2008, \apjl, 688,
  L91

\bibitem[{Maund \& Smartt(2005)}]{Maund:2005p492}
Maund, J.~R. \& Smartt, S.~J. 2005, MNRAS, 360, 288

\bibitem[{{Maund} \& {Smartt}(2009)}]{2009Sci...324..486M}
{Maund}, J.~R. \& {Smartt}, S.~J. 2009, Science, 324, 486

\bibitem[{Maund {et~al.}(2005)Maund, Smartt, \& Danziger}]{Maund:2005p490}
Maund, J.~R., Smartt, S.~J., \& Danziger, I.~J. 2005, MNRAS: Letters, 364, L33

\bibitem[{{Meikle} {et~al.}(2006){Meikle}, {Mattila}, {Gerardy}, {Kotak},
  {Pozzo}, {van Dyk}, {Farrah}, {Fesen}, {Filippenko}, {Fransson}, {Lundqvist},
  {Sollerman}, \& {Wheeler}}]{2006ApJ...649..332M}
{Meikle}, W.~P.~S., {Mattila}, S., {Gerardy}, C.~L., {et~al.} 2006, \apj, 649,
  332

\bibitem[{Misra {et~al.}(2007)Misra, Pooley, Chandra, Bhattacharya, Ray, Sagar,
  \& Lewin}]{Misra:2007p1117}
Misra, K., Pooley, D., Chandra, P., {et~al.} 2007, MNRAS, 381, 280

\bibitem[{Munari \& Zwitter(1997)}]{Munari:1997p1098}
Munari, U. \& Zwitter, T. 1997, A\&A, 318, 269

\bibitem[{Nakano \& Itagaki(2006)}]{Nakano:2006p559}
Nakano, S. \& Itagaki, K. 2006, IAU Circ., 8773, 1

\bibitem[{Nakano {et~al.}(2006)Nakano, Itagaki, \& Kadota}]{Nakano:2006p1928}
Nakano, S., Itagaki, K., \& Kadota, K. 2006, CBET, 756, 1

\bibitem[{Pilyugin {et~al.}(2004)Pilyugin, V{\'\i}lchez, \&
  Contini}]{Pilyugin:2004p1923}
Pilyugin, L.~S., V{\'\i}lchez, J.~M., \& Contini, T. 2004, A\&A, 425, 849

\bibitem[{Poelarends {et~al.}(2008)Poelarends, Herwig, Langer, \&
  Heger}]{Poelarends:2008p103}
Poelarends, A. J.~T., Herwig, F., Langer, N., \& Heger, A. 2008, ApJ, 675, 614

\bibitem[{{Prieto} {et~al.}(2008){Prieto}, {Kistler}, {Thompson}, {Y{\"u}ksel},
  {Kochanek}, {Stanek}, {Beacom}, {Martini}, {Pasquali}, \&
  {Bechtold}}]{2008ApJ...681L...9P}
{Prieto}, J.~L., {Kistler}, M.~D., {Thompson}, T.~A., {et~al.} 2008, \apjl,
  681, L9

\bibitem[{Sahu {et~al.}(2006)Sahu, Anupama, Srividya, \&
  Muneer}]{Sahu:2006p1119}
Sahu, D.~K., Anupama, G.~C., Srividya, S., \& Muneer, S. 2006, MNRAS, 372, 1315

\bibitem[{Schlegel {et~al.}(1998)Schlegel, Finkbeiner, \&
  Davis}]{Schlegel:1998p1933}
Schlegel, D.~J., Finkbeiner, D.~P., \& Davis, M. 1998, ApJ v.500, 500, 525

\bibitem[{Schoeniger \& Sofue(1997)}]{Schoeniger:1997p1926}
Schoeniger, F. \& Sofue, Y. 1997, A\&A, 323, 14

\bibitem[{Siess(2007)}]{Siess:2007p64}
Siess, L. 2007, A\&A, 476, 893

\bibitem[{{Smartt}(2009)}]{2009ARA&A..47...63S}
{Smartt}, S.~J. 2009, \araa, 47, 63

\bibitem[{{Smartt} {et~al.}(2009){Smartt}, {Eldridge}, {Crockett}, \&
  {Maund}}]{2009MNRAS.395.1409S}
{Smartt}, S.~J., {Eldridge}, J.~J., {Crockett}, R.~M., \& {Maund}, J.~R. 2009,
  \mnras, 395, 1409

\bibitem[{{Smartt} {et~al.}(2001){Smartt}, {Gilmore}, {Trentham}, {Tout}, \&
  {Frayn}}]{2001ApJ...556L..29S}
{Smartt}, S.~J., {Gilmore}, G.~F., {Trentham}, N., {Tout}, C.~A., \& {Frayn},
  C.~M. 2001, \apjl, 556, L29

\bibitem[{Smartt {et~al.}(2003)Smartt, Maund, Gilmore, Tout, Kilkenny, \&
  Benetti}]{Smartt:2003p1939}
Smartt, S.~J., Maund, J.~R., Gilmore, G.~F., {et~al.} 2003, MNRAS, 343, 735

\bibitem[{Smartt {et~al.}(2004)Smartt, Maund, Hendry, Tout, Gilmore, Mattila,
  \& Benn}]{Smartt:2004p444}
Smartt, S.~J., Maund, J.~R., Hendry, M.~A., {et~al.} 2004, Science, 303, 499

\bibitem[{Solanes {et~al.}(2002)Solanes, Sanchis, Salvador-Sol{\'e},
  Giovanelli, \& Haynes}]{Solanes:2002p1225}
Solanes, J.~M., Sanchis, T., Salvador-Sol{\'e}, E., Giovanelli, R., \& Haynes,
  M.~P. 2002, AJ, 124, 2440

\bibitem[{Stanishev \& Nielsen(2006)}]{Stanishev:2006p1929}
Stanishev, V. \& Nielsen, T.~B. 2006, CBET, 737, 1

\bibitem[{{Tully}(1988)}]{1988ngc..book.....T}
{Tully}, R.~B. 1988, {Nearby Galaxies Catalog} (Cambridge University Press,
  1988, 221 pages.)

\bibitem[{{Van Dyk} {et~al.}(2003{\natexlab{a}}){Van Dyk}, {Li}, \&
  {Filippenko}}]{2003PASP..115..448V}
{Van Dyk}, S.~D., {Li}, W., \& {Filippenko}, A.~V. 2003{\natexlab{a}}, \pasp,
  115, 448

\bibitem[{{Van Dyk} {et~al.}(2003{\natexlab{b}}){Van Dyk}, Li, \&
  Filippenko}]{VanDyk:2003p487}
{Van Dyk}, S.~D., Li, W., \& Filippenko, A.~V. 2003{\natexlab{b}}, PASP, 115,
  1289

\bibitem[{{Van Dyk} {et~al.}(2003{\natexlab{c}}){Van Dyk}, Li, \&
  Filippenko}]{VanDyk:2003p1940}
{Van Dyk}, S.~D., Li, W., \& Filippenko, A.~V. 2003{\natexlab{c}}, PASP, 115, 1

\bibitem[{{van Loon} {et~al.}(2005){van Loon}, Cioni, Zijlstra, \&
  Loup}]{vanLoon:2005p1938}
{van Loon}, J.~T., Cioni, M.-R.~L., Zijlstra, A.~A., \& Loup, C. 2005, A\&A,
  438, 273

\bibitem[{Zwitter {et~al.}(2004)Zwitter, Munari, \&
  Moretti}]{Zwitter:2004p1930}
Zwitter, T., Munari, U., \& Moretti, S. 2004, IAU Circ., 8413, 1

\end{thebibliography}

\newpage

\appendix
\section{A Note on Extinction}
\label{apend:ext_note}
 
In all cases in this paper, we have used our best estimate for the extinction towards each SN as a proxy for that of its progenitor star.  We therefore assume that the extinction towards each SN is the same as that towards its progenitor before explosion.  However, if the progenitor stars happened to be surrounded by a significant quantity of circumstellar dust, this might not be true \citep[e.g.,][]{2009MNRAS.398.1041B}.  Dust within a certain radius could be photo-evaporated by the SN explosion \citep{Dwek:1983p1936}, effectively reducing the extinction measured towards a SN compared with that towards its progenitor star.  \citet{2009MNRAS.398.1041B} have estimated that it is possible for the UV-optical luminosity of a type II-P SN to destroy circumstellar dust equating to several tens of magnitudes of optical extinction.

All this is of little concern if the progenitor star is free of such circumstellar material, or if this material is far enough from the star to avoid evaporation \citep[e.g.,][]{2006ApJ...649..332M}.  If a progenitor star was detected across a sufficient number and range of band-passes one could directly measure its extinction.  Such direct measurement could reveal the presence of circumstellar dust that was subsequently destroyed by the SN.  Unfortunately none of the three SNe studied in this paper fulfil these requirements. 

In $\S$6.2 of \citet{2009MNRAS.395.1409S}, we have discussed the progenitor extinction problem with relation to 20 type II-P SNe, including the three featured in this paper.  We refer the reader to \citet{2009MNRAS.395.1409S} for the detailed discussion, but highlight some of the main points here.  The distribution of extinction values estimated for our progenitor sample was more or less consistent with that for optically selected red supergiants in the LMC and SMC \citep{Levesque:2006p1920}, albeit with some evidence to suggest there were greater numbers of objects from our sample in the lowest extinction bin than would be typical for red supergiants \citep[see Figure 5][]{2009MNRAS.395.1409S}.  An arbitrary increase in $A_V$ of 0.3 mag for the 5 lowest reddening values in our sample brings the mean into line with that of the LMC/SMC red supergiants.  The progenitors of SNe 2006my and 2006ov are two of the five objects ostensibly suffering the lowest extinctions.  Applying the above arbitrary increase raises the upper mass limits for these two progenitors to: 2006my\,$<$\,14M$_{\odot}$; 2006ov\,$<$\,11M$_{\odot}$.  

Highly reddened ($A_V \sim 10$) red supergiants do exist \citep[e.g.][]{vanLoon:2005p1938} and these, being optically obscured, would not be included in the sample of \citet{Levesque:2006p1920}.  Our comparison sample is therefore biased towards lower extinctions, although to what extent is unclear since the relative number of heavily extinguished to low/moderately extinguished red supergiants is unknown.  Without these relative numbers, we favour the progenitors to have come from a population of low/moderately extinguished red supergiants, although ultimately we cannot rule out that any of the three progenitors studied in this paper could have experienced much higher reddening than estimated.

How would this essentially arbitrary source of extinction affect our results?  Assuming arbitrary extinction, no constraints whatsoever can be placed on the mass limits or spectral types of the progenitors of SNe 2006my and 2006ov, since in both cases we do not detect a progenitor star.  In the case of SN 2004et some constraints are possible.  Assuming the progenitor photometry of $\S$\ref{sec:04et_final_prog_properties}, the colour of any fitted progenitor model must satisfy the observed (apparent) colour range of \mbox{$(R\!-\!I)$}=1.51$\pm$0.31.  For a progenitor of given spectral type, a range of extinctions between minimum and maximum values will satisfy this condition.  Hotter/bluer spectral types will require higher minimum extinctions, while such increases in extinction will rule out cooler spectral types.  The result is that, as extinction is increased, we can fit progenitors of higher luminosity/mass, but these must also be of earlier spectral type.

\begin{figure}
    \centering
    \includegraphics[width=80mm]{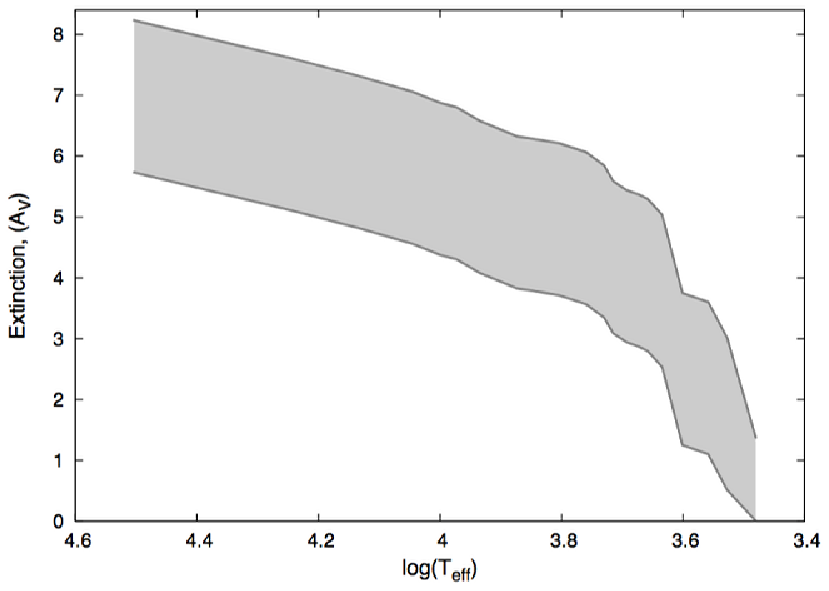}
    \includegraphics[width=80mm]{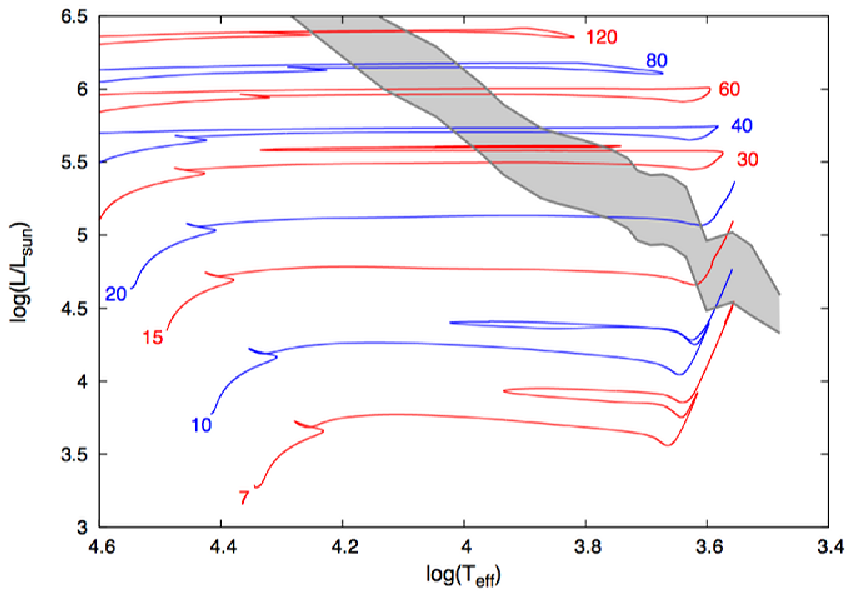}
    \caption{The properties of the progenitor of SN 2004et, assuming arbitrary levels of extinction. {\bf Top panel:} Range of extinction levels required to fit a star of certain spectral type to the progenitor photometry.  Increased extinction can lead to higher luminosity/mass progenitors, but such stars are also of earlier/hotter spectral type.  {\bf Bottom panel:} The shaded area denotes the range of luminosities for a progenitor of a given spectral type/effective temperature.}
\label{fig:red_ext}
\end{figure}

The results of a simple calculation using the photometry of the SN 2004et progenitor, as derived in $\S$\ref{sec:04et_final_prog_properties}, and the intrinsic \mbox{$(R\!-\!I)_o$} colours of supergiant stars from \citet{2000asqu.book..381D} are plotted in Figure~\ref{fig:red_ext}.  The minimum and maximum values of extinction, $A_V$, required to fit each spectral type are easily calculated from:

\begin{equation}
A_V=(A_R\!-\!A_I)/0.272
\label{eqn:min_ext}
\end{equation}

\noindent where,

\begin{equation}
(A_R\!-\!A_I)=(R\!-\!I)-(R\!-\!I)_o
\end{equation}

The denominator of eqn.~\ref{eqn:min_ext} comes from the ratios $A_R/A_V$=0.751 and $A_I/A_V$=0.479 from \citet{Cardelli:1989p1230}.  The minimum and maximum luminosities plotted in Figure~\ref{fig:red_ext} were calculated from the I-band progenitor photometry ($\S$\ref{sec:04et_final_prog_properties}), the distance modulus ($\S$\ref{sec:04et_mass}), the bolometric and colour corrections from \citet{2000asqu.book..381D}, and the minimum and maximum extinction values derived for each spectral type.  {\sc stars} stellar models for stars of initial masses between 7 - 120 M$_{\odot}$ are also plotted.  These can be used as a rough guide to the initial stellar mass for a progenitor of given spectral type.  Alternatively, the plotted luminosities can be compared directly with model endpoint luminosities in Figure~\ref{fig:STARS}.  

In the introduction to this paper we discussed the X-ray and radio detections of SN~2004et \citep{Argo:2005p1792,Misra:2007p1117,MartiVidal:2007p1547}, while in $\S$\ref{sec:echo} we show a NIR re-brightening some 4 years post-explosion and discuss the results of Kotak et al. (2009) who find evidence of ejecta-CSM interaction.  These observations infer the presence of {\em significant} amounts of circumstellar material (CSM) around the progenitor star.  One might therefore argue, with some justification, that the progenitor of SN 2004et was significantly extinguished by CSM, some of which may have been destroyed in the SN explosion.  Our simple analysis suggests that if the extinction was greater than $A_V\approx4.0$, the progenitor star could not have been a red supergiant.  On the other hand, if it was a red supergiant it could not have had an initial mass greater than $\sim$16 M$_{\odot}$.

It should be noted that to assume a progenitor of earlier spectral type automatically requires it to be more compact.  For example a G0 supergiant has a radius of R$_{\ast}\approx100$R$_{\odot}$.  Such radii are not easy to reconcile with the fact that SN~2004et was a typical II-P \citep{Misra:2007p1117,Sahu:2006p1119,2010MNRAS.404..981M}.  SNe II-P require larger radii of the order 400-1000R$_{\odot}$ \citep{1976ApJ...207..872C,Arnett:1980p1887}, which are typical for red supergiant stars.

\label{lastpage}
\end{document}